\documentclass{aastex631}

\renewcommand{\ion}[2]{#1\,\textsc{#2}}

\newcommand{\hinode}{\textit{Hinode}}

\usepackage{bold-extra}

\shorttitle{Additions to the Spectrum of \ion{Fe}{ix}}
\shortauthors{Ryabtsev et al.}
\graphicspath{{./}{figures/}}

\begin{document}

\title{Additions to the Spectrum of Fe\,IX in the 110--200~\AA\ Region}

\correspondingauthor{Peter R. Young}
\email{peter.r.young@nasa.gov}

\author[0000-0002-5321-5406]{Alexander N. Ryabtsev}
\affiliation{Institute of Spectroscopy, Russian Academy of Sciences, Troitsk, Moscow 108840, Russia}

\author{Edward Y. Kononov}
\affiliation{Institute of Spectroscopy, Russian Academy of Sciences, Troitsk, Moscow 108840, Russia}

\author[0000-0001-9034-2925]{Peter R. Young}
\affiliation{NASA Goddard Space Flight Center, Greenbelt, MD 20771, USA}
\affiliation{Northumbria University, Newcastle upon Tyne, NE1 8ST, UK}



\begin{abstract}

The spectrum of eight-times ionized iron, \ion{Fe}{ix}, was studied in the 110--200 \AA\ region. A low inductance vacuum spark and a 3-m grazing incidence spectrograph were used for the excitation and recording of the spectrum. Previous analyses of \ion{Fe}{ix} have been greatly extended and partly revised. The number of known lines in the $3p^53d$--$3p^54f$ and $3p^53d$--$3p^43d^2$ transition arrays is extended to 25 and 81, respectively. 
Most of the identifications of the \ion{Fe}{ix} lines from the $3p^53d$--$3p^43d^2$ transition array in the solar spectrum have been confirmed and several new identifications are suggested.

\end{abstract}



\section{Introduction} \label{sec:intro}

Eight-times ionized iron, \ion{Fe}{ix}, belongs to the argon isoelectronic sequence with a fully-occupied $3p^6$ ground configuration. The strongest line in the \ion{Fe}{ix} spectrum is the $3p^6$ $^1S_0$--$3p^53d$ $^1P_1$ resonance transition at 171.073~\AA, which was identified independently by \citet{1965Natur.206..390G} and \citet{1965Natur.206..176A}. The reference wavelength for the line listed in the NIST database \citep{kramida20} comes from the solar spectral atlas of \citet{1972ApJ...175..493B}. In the past four decades the 171~\AA\ line has been a  popular choice for solar extreme ultraviolet (EUV) imaging instruments on account of its strength, relative isolation in the spectrum, and the proximity of an aluminium absorption edge at 170~\AA. EUV imaging instruments make use of multilayer coatings to yield narrow bandpasses that are typically 10\%\ of the central wavelength. However, the high density of lines in the EUV means that multiple strong lines can contribute to the imaging channel emission. Aluminium filters are used to block visible radiation in these instruments, and the aluminium edge at 170~\AA\  strongly attenuates emission to the short wavelength side of \ion{Fe}{ix} 171.07~\AA, yielding a relatively clean bandpass. Because of this a filter at or close to 171~\AA\ has been used for several telescopes on spacecraft, including the \textit{EUV Imaging Telescope} \citep[EIT:][]{1995SoPh..162..291D}, the \textit{Transition Region and Coronal Explorer} \citep[TRACE:][]{1999SoPh..187..229H}, the \textit{Extreme Ultraviolet Imager} \citep[EUVI:][]{2008SSRv..136...67H}, the \textit{Atmospheric Imaging Assembly} \citep[AIA:][]{2012SoPh..275...17L}, and most recently the \textit{Extreme-Ultraviolet Imager} \citep[EUI:][]{2020A+A...642A...8R} on \textit{Solar Orbiter}. The 171.07~\AA\ line has also been measured in spectra of the cool stars Procyon \citep{1995ApJ...443..393D} and $\alpha$ Centauri \citep{1997ApJ...478..403D}.

Other \ion{Fe}{ix} lines in the EUV are much weaker, but the ratio of the lines at 241.74 and 244.91~\AA\ (decays from the $3p^53d$ $^3P_{2,1}$  states to the ground) has long been recognised as a useful density diagnostic of the solar atmosphere \citep{1978ApJ...219..304F}. 

The first excited configuration of \ion{Fe}{ix}, $3s^23p^53d$, has 12 fine-structure levels and yields a rich spectrum of forbidden lines between 1738~\AA\ and 2.9~$\mu$m. Many of these lines have been measured during solar eclipses, and a summary is provided by \citet{2018ApJ...852...52D}.

At soft X-ray wavelengths, there are three groups of lines coming from $n=4$ configurations. The $3p^6$--$3p^54d$ and $3p^6$--$3p^54s$ transition arrays give two pairs of lines at 82.43 and 83.46~\AA, and 103.57 and 105.21~\AA, respectively, and these lines were reported in a solar spectrum by \citet{1973ApJ...181.1009M}. The $3p^53d$--$3p^54f$ array gives a group of lines between 111 and 120~\AA\ and an additional pair of lines at 133.92 and 136.57~\AA. None of these lines have been reported from solar spectra, but they have been studied with laboratory spectra. This array is the subject of study in the present article and is discussed in Section~\ref{sect.4f}.



Recent advances in the spectral analysis of \ion{Fe}{ix} have largely been motivated by the solar spectra obtained with the \textit{Extreme ultraviolet (EUV) Imaging Spectrometer} \citep[EIS:][]{2007SoPh..243...19C} on board the \textit{Hinode} spacecraft, launched in 2006. EIS has a spectral resolution of 3000 and it covers the wavelength ranges 170--212 and 246--292~\AA, which contain a wealth of emission lines from iron ions, ranging from \ion{Fe}{vii} up to \ion{Fe}{xxiv} \citep{2007PASJ...59S.857Y}. \citet{2008ApJS..176..511B} reported that about half of the lines in the EIS wavelength ranges were unidentified. \citet{2009ApJ...691L..77Y} was the first to report new \ion{Fe}{ix} identifications in the EIS wavelength bands, and further studies of the \ion{Fe}{ix} spectrum have been performed by \citet{2009A+A...508..501D} \citet{2009ApJ...706....1L,2009ApJ...707.1191L} \citet{2009ApJ...707..173Y}, \citet{2012A+A...537A..22O} and \citet{2014A+A...565A..77D}. Laboratory studies of the \ion{Fe}{ix} spectrum in the EUV have been performed by \citet{2009ApJ...696.2275L}, \citet{2012ApJS..201...28B} and \citet{2018ApJ...854..114B}. These works have resulted in a number of line identifications from the $3p^53d$--$3p^43d^2$ transitions and around two dozen lines suggested as due to \ion{Fe}{ix} but without definite identifications. 



\begin{figure}[t]
    \centering
    \includegraphics[width=\textwidth]{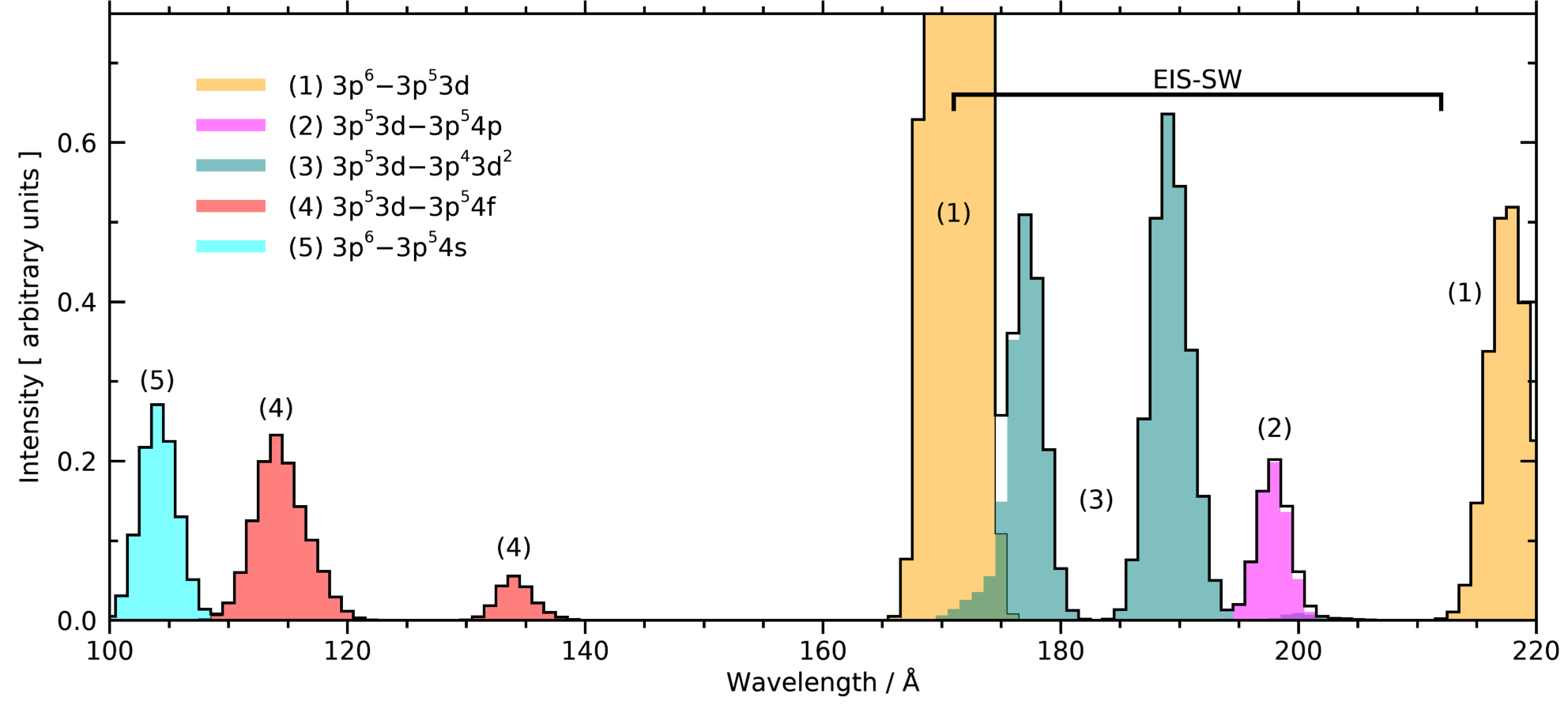}
    \caption{A \textsf{CHIANTI} synthetic spectrum showing the location of known \ion{Fe}{ix} lines. Colors and numbers denote different transition arrays. Line profiles are represented as Gaussians with a full-width at half-maximum of 3~\AA. A temperature of 0.8~MK and an electron number density of $4.0\times 10^8$~cm$^{-3}$ were used. The wavelength range of the EIS short-wavelength (SW) band is indicated.}
    \label{fig.spec}
\end{figure}

The present work focuses on line identifications in the 110--200~\AA\ range from the $3p^53d$--$3p^54f$  and $3p^53d$--$3p^43d^2$ transition arrays. Figure~\ref{fig.spec} gives an overview of the known (prior to the present work) \ion{Fe}{ix} transitions in the 100--220~\AA\ as found in version~10 of the \textsf{CHIANTI} atomic database \citep{2016JPhB...49g4009Y,2021ApJ...909...38D}. The spectrum was generated using a temperature of 0.8~MK and an electron number density of $4.0\times 10^8$~cm$^{-3}$, which are typical of conditions in the solar atmosphere. Emission line widths have been set to 3~\AA\ in order to group nearby lines together. Five transition arrays are highlighted: the two studied in this work, and the $3p^6$--$3p^54s$, $3p^53d$--$3p^54p$ and $3p^6$--$3p^53d$ arrays. The latter yields the 171~\AA\ line, which towers over the other features to an intensity peak of 10 on the plot scale. The only known line from the $3p^53d$--$3p^54p$ array is at 197.86~\AA\ \citep{2009ApJ...691L..77Y}. Note that there are many more transitions in \textsf{CHIANTI} in this part of the spectrum that are not shown because they only have theoretical wavelengths.

Our new laboratory study uses spectra excited in a vacuum spark and recorded with a higher resolution than in all previous papers. The number of known lines in the $3p^53d$--$3p^54f$ transition array is extended to 25. Several previous identifications were corrected resulting in changes to the energies of the corresponding levels. Eighty-one lines of the $3p^53d$--$3p^43d^2$ array have been identified.  Most of the previous identifications from this array in the solar spectrum have been  confirmed and several new ones have been made. Lines previously assigned to \ion{Fe}{ix} but not identified have been classified.
This article continues a series of publications on laboratory high resolution studies of iron ion spectra \citep{2017EPJWC.13203043R,2021ApJ...908..104Y,2022ApJS..258...37K} relevant for diagnostics of solar plasma.

Section~\ref{sect.expt} describes the experimental setup used to obtain the \ion{Fe}{ix} spectra. Section~\ref{sect.results} gives the line identification results for the two transition arrays, and we give our conclusions in Section~\ref{sect.conc}.

\section{Experiment}\label{sect.expt}

The procedures used in our earlier studies of the iron ions \citep[see, for example,][]{2022ApJS..258...37K} were followed in the present measurements. In short, the spectra were taken with a 3~m grazing incidence $(5^\circ)$ spectrograph equipped with a 3600 line~mm$^{-1}$ grating. The spectra were excited in a three-electrode vacuum spark run with peak currents up to 100 kA. The high current tracks on the ORWO UV-2 photographic plates taken for an  analysis of \ion{Fe}{viii} \citep{1980OptSp..48..348R} were measured. The tracks were scanned on an EPSON EXPRESSION scanner and then digitized and measured using the \textsf{Gfit} code \citep{engstrom98}. 


\begin{figure}
    \centering
    \includegraphics[scale=1.4, angle=90]{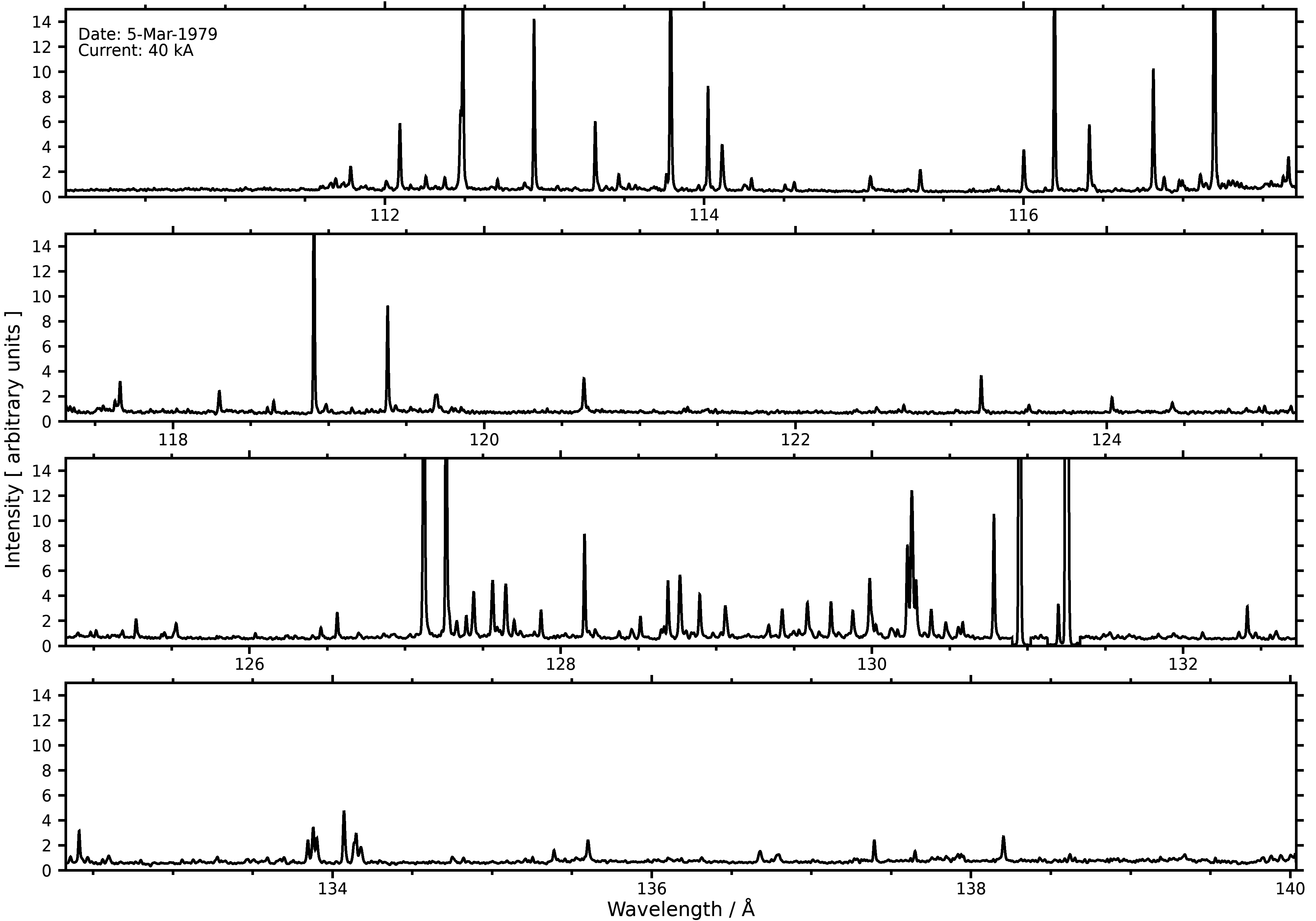}
    \caption{The photoplate spectrum  obtained at a current of 40~kA.}
    \label{fig.specs1}
\end{figure}


\begin{figure}
    \centering
    \includegraphics[scale=1.4, angle=90]{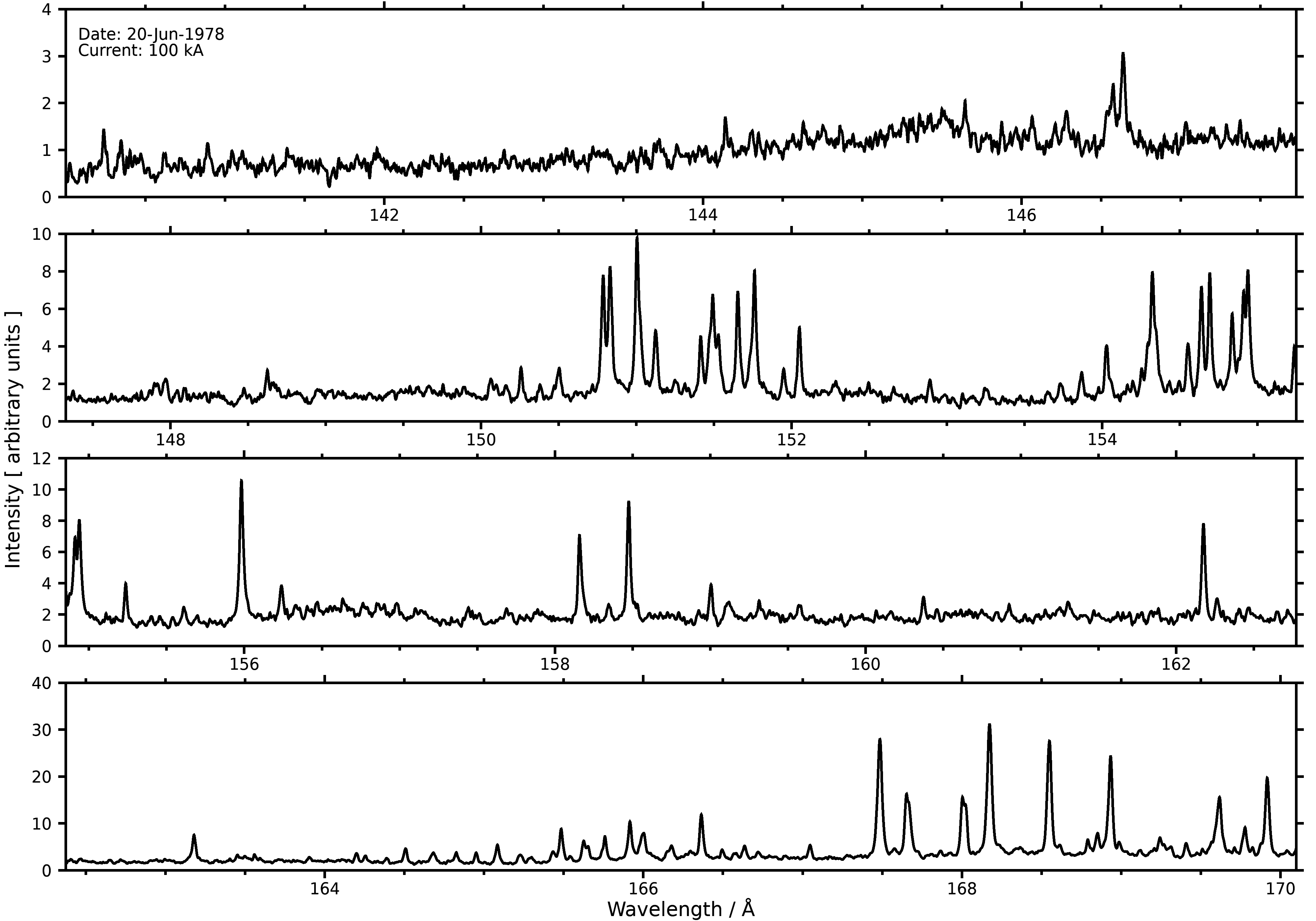}
    \caption{A section of the photoplate spectrum obtained at a current of 100~kA.}
    \label{fig.specl1a}
\end{figure}

\begin{figure}
    \centering
    \includegraphics[scale=1.4, angle=90]{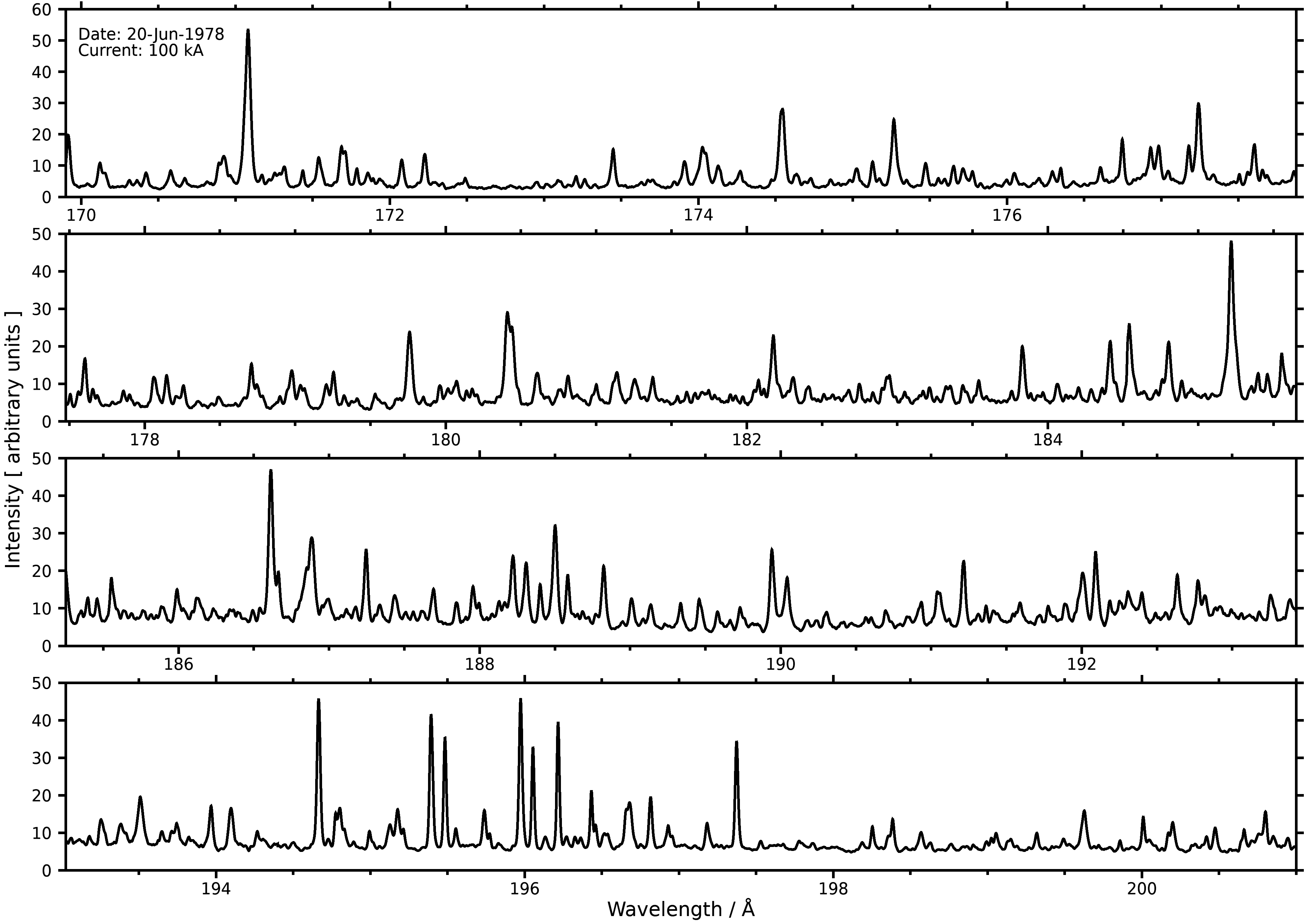}
    \caption{A section of the photoplate spectrum obtained at a current of 100~kA.}
    \label{fig.specl1b}
\end{figure}

Spectra were also recorded on phosphor imaging plates (Fuji BAS-TR) \citep{2017EPJWC.13203043R}. These spectra were scanned using a 0.01~mm scanning step with a Typhoon FLA 9500 reader and processed and analysed with the ImageQuant TL 7.0 image analysis software. For final reduction of the spectra  the \textsf{GFit} code again was employed. Because of a high level of background, the imaging plate spectra taken at high currents were useful only for wavelengths above  around 140~\AA.

Figure~\ref{fig.specs1} shows the photoplate spectrum between 110 and 140~\AA\ obtained for a current of 40~kA, and Figures~\ref{fig.specl1a} and \ref{fig.specl1b} show the photoplate spectrum between 140 and 201~\AA\ obtained at 100~kA. The wavelength scales are not the final ones used to derive the \ion{Fe}{ix} wavelengths as a grating formula with averaged parameters was used. The final \ion{Fe}{ix} wavelengths, given in Tables~\ref{tbl.1} and \ref{tbl.3}, were obtained following a reduction with reference lines and a correction curve of the photoplate. Close-ups of selected portions of the spectra are shown in Figures~\ref{fig.2}--\ref{fig.3}, where spectra obtained for different currents are compared.

The iron line wavelengths were measured by first taking  photoplate spectra with an iron anode and a titanium cathode of the spark, giving a mixed spectrum of iron and titanium lines. The known titanium wavelengths \citep{svensson69} were used to derive reference wavelengths for those iron lines that are not blended with titanium lines. These iron line wavelengths were then used as secondary standards when reducing the pure iron spectra obtained with both iron electrodes in the vacuum spark.
The width of the \ion{Fe}{ix} lines on the photoplate spectra in the region above 140~\AA\ was about 0.03 \AA\, whereas on the imaging plate spectra the width was larger by about 20\%. 
The photoplate containing the spectrum below 140~\AA\ that was used for the analysis of the $3p^53d$--$3p^54f$ transitions had a better resolution, with line widths around 0.02~\AA. 
The photoplate spectra possessed two other advantages with respect to the imaging plate spectra that were important for the wavelength measurements. Firstly, the signal-to-noise was better due to a smaller background, and secondly, the line shape was smoother due to the approximately two times smaller scanning step of the EPSON EXPRESSION scanner in comparison with the Typhoon FLA 9500 reader. The root-mean-square deviation of the reference titanium, as well as the secondary iron, lines from the calibration curve was 0.002~\AA, and the deviations from the mean values of the wavelengths obtained from the measurements of different spectra recordings was below 0.001~\AA. Because \citet{svensson69}  claimed an uncertainty of 0.004~\AA\ for their wavelengths we adopt this value as our wavelength uncertainty for single, unperturbed lines. 
This estimation is supported by the deviations of the measured wavelengths from the Ritz values calculated in the present spectrum analysis.
Where possible,  the line intensities were obtained from the imaging plate spectra due to the greater linearity of the response. The intensities will be given on an  arbitrary scale without taking into account the wavelength dependence of the spectrograph efficiency and recording media sensitivity.

Based on our past experience with the vacuum spark, the plasma is in approximate local thermodynamic equilibrium with an effective electron temperature. Rough estimates following the procedure of \citet{2022ApJS..258...37K} for \ion{Fe}{vii} give a value of 10~eV. It is not possible to estimate a density from the spectra, but we expect a value around $10^{17}$~cm$^{-3}$ based on measurements from  similar plasma discharges. The density would be expected to vary as the current is varied to modify the charge balance.

The \ion{Fe}{ix} analysis was guided by calculations with the Cowan atomic code \citep{1981tass.book.....C,kramida19}. The set of even parity configurations used was: $3p^6$, $3p^5np$ ($n=4$ to 6), $3p^5nf$ ($n=4$ to 6), $3p^43d^2$, $3p^44s^2$,  $3p^43d4s$, $3p^44d4s$, $3s3p^63d$ and $3s3p^64s$. The set of interacting odd parity configurations was restricted to $3p^5nd$ ($n=3$ to 6), $3p^54s$, $3p^55s$, $3s3p^64p$ and $3s3p^64f$. The calculated energy levels and transition probabilities were used as input to the  \textsf{IDEN2} line identification code \citep{2018CoPhC.225..149A}.

\section{Results}\label{sect.results}

\subsection{The $3p^53d$--$3p^54f$ transitions}\label{sect.4f}

Laboratory measurements of the $3p^53d$--$3p^54f$ transition array were performed in the 1970s, with \citet{1971ApJ...166..683W} listing 12 lines in the 111--117~\AA\ range. Nine of these were  measured by \citet{1972JPhB....5.1255F} with an accuracy of $\pm 0.01$~\AA, and further work was done by \citet{1976JOSA...66..240S} who remeasured all 12 lines, yielding energies for 10 of the 12 $3p^54f$ levels  with an estimated uncertainty of $\pm200$~cm$^{-1}$. 

More recently, \citet{2012A+A...537A..22O}  provided a list of  20 identified $3p^53d$--$3p^54f$ lines and all level energies of the $3p^54f$ configuration. The wavelengths include the 11 values from \citet{1972JPhB....5.1255F}, but the source of the other nine wavelengths is not given. We speculate that the authors had access to the original plate of Fawcett and performed new measurements as done in the earlier works of \citet{2009A+A...508..501D} and \citet{2012A+A...546A..97D}. \citet{2012A+A...537A..22O} did not provide uncertainties on their derived level energies.



Our results for this transition array are displayed in Table~\ref{tbl.1} for the wavelengths and in Table~\ref{tbl.2} for the energy levels of the $3p^54f$ configuration. Many weak lines were added to the previous analyses of these transitions extending to 25 the number of identified lines and resulting in the location of all the $3p^54f$ levels with uncertainties 30~cm$^{-1}$ or less. The wavelengths of several lines marked by VII in Table~\ref{tbl.1} are close to the \ion{Fe}{vii} wavelengths of \citet{1981PhyS...23....7E}. Indeed, these lines together with the other \ion{Fe}{vii} lines are present in our tracks with ``colder" iron spectra but their influence on the intensities and wavelengths  of the \ion{Fe}{ix} lines is negligible in our case. One line with Ritz value 112.463(3)~\AA\ is not measured, possibly being masked by the \ion{Fe}{viii} 112.472~\AA\ line.

The relative intensities measured in the photographic plate spectrum are given on an arbitrary linear scale. The 250 value was adopted for the strongest line (113.789~\AA) for convenient comparison with the corresponding $gA$ values. ($g$ is the statistical weight of the transition upper level and $A$ the radiative decay rate.) The wavelength responses of the grating and photoplate were not taken into account. Even for a narrow wavelength range, the intensities should be considered as more qualitative than quantitative because of a possible error in the model characteristic curve of the photoplate used for a transformation of the measured line densities to their intensities.

\begin{figure}
    \centering
    \includegraphics[width=0.8\textwidth]{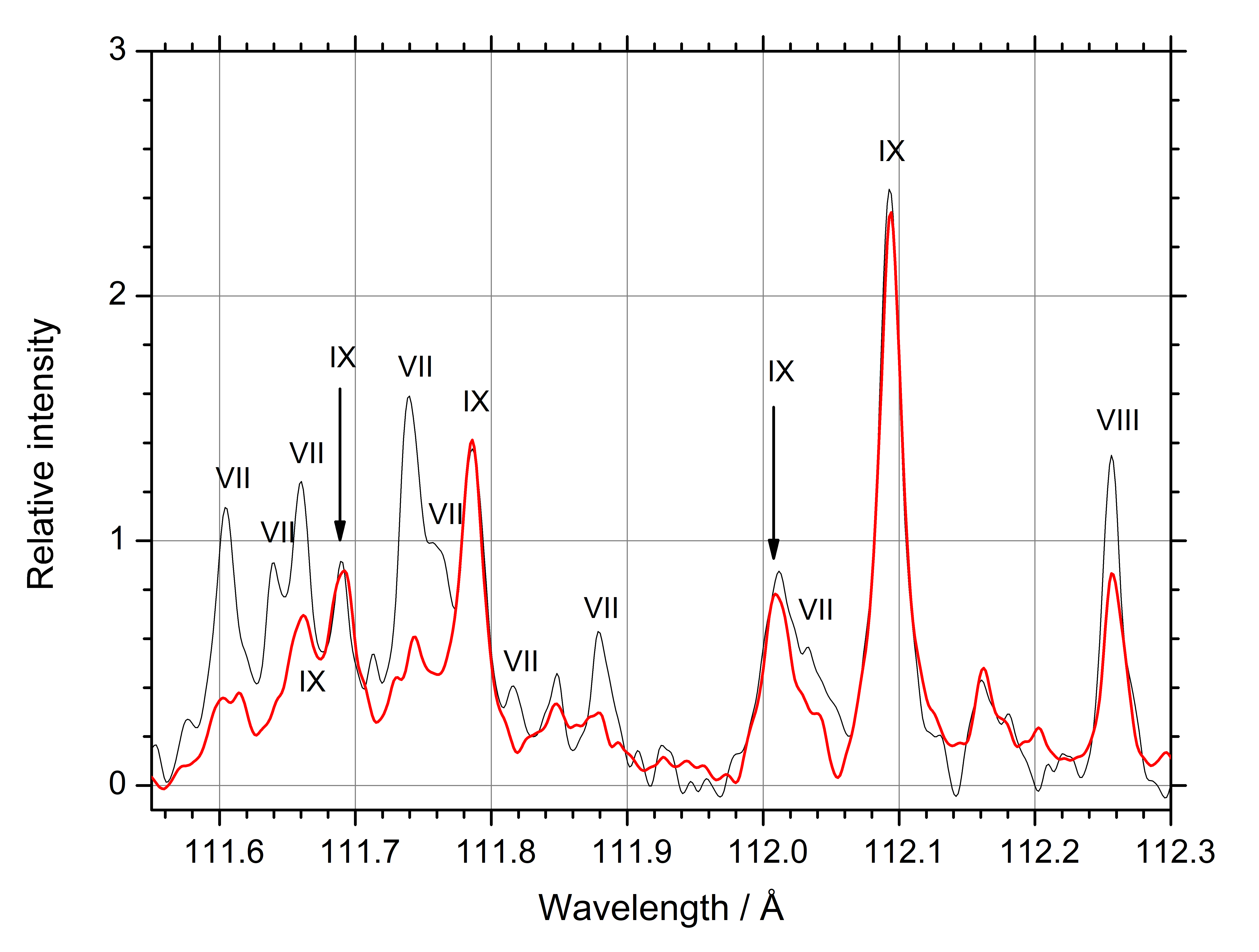}
    \caption{Two iron spectra correspond to ``hot" (red) and ``cold" (black) excitation conditions. The lines are marked by the symbols: \textsc{vii} -- \ion{Fe}{vii}, \textsc{viii} -- \ion{Fe}{viii} and \textsc{ix} -- \ion{Fe}{ix}. Lines due to the three iron species are distinguished by their relative intensities in the two spectra. Arrows indicate the \ion{Fe}{ix} lines at 111.692 and 112.011~\AA.}
    \label{fig.2}
\end{figure}

Table 1 contains a comparison with the previous measurements by \citet{1976JOSA...66..240S} and \citet{2012A+A...537A..22O}. Good agreement of all measurements with some exceptions is seen. The two lines at 111.713 and 112.031~\AA\ were listed by \citet{2012A+A...537A..22O} as the transitions from the $3p^54f$ $^3D_1$ level. A part of the iron spectrum in the region of these lines, taken at two modes of the spark operation, is shown in Figure 2. 
\ion{Fe}{ix} lines are clearly distinguished from those of \ion{Fe}{vii} and \ion{Fe}{viii} by the change in intensity  from ``cold" to ``hot" spark conditions. The \ion{Fe}{ix} lines have about the same intensity in the displayed spectra, while the lower ionization states have reduced intensities in the ``hot" spectrum.
Both the 111.713 and 112.031~\AA\ lines identified by \citet{2012A+A...537A..22O} are seen to belong to lower ionization states than \ion{Fe}{ix}. The feature at 112.031~\AA\ is a known line of \ion{Fe}{vii}, measured at 112.030~\AA\ by  \citet{1981PhyS...23....7E}.
The lines at 111.692 and 112.011~\AA\ (marked with arrows on Figure~\ref{fig.2}) show behaviour similar to the other \ion{Fe}{ix} lines and were adopted as the transitions from the $3p^54f$ $^3D_1$ level. The wavelengths are in good agreement with those of \citet{1976JOSA...66..240S}. 

The line of 118.27~\AA\ identified by \citet{1976JOSA...66..240S} as the $3p^53d$ $^1F_3$ -- $3p^54f$ $^3F_4$ transition has a low transition probability and is absent in our spectrum. Its Ritz value is 118.220(3)~\AA.

Previous \ion{Fe}{ix} studies did not provide definitive line identifications for the $3p^54f$ $^1D_2$ and $3p^54f$ $^3F_2$ levels.
Swartz et al.\ tentatively suggested that the two lines at 134.743 and 115.46~\AA\ belong to decays from the $3p^54f$ $^1D_2$ level to the $3p^53d$ $^1P_1$ and $^3D_1$ levels, resulting in a value of 1\,326\,700~cm$^{-1}$ for the $^1D_2$ level. It should be noted that both lines are not present in our ``hot" spectrum, the second one being possibly the 115.472~\AA\ line of \ion{Fe}{vii} \citep{1981PhyS...23....7E}.  \citet{2002ApJ...578..648L} recorded an \ion{Fe}{ix} spectrum at the Lawrence Livermore electron beam ion trap (EBIT).
The spectral resolution was relatively low (around 300 at 100~\AA) and the authors measured two weak lines at 134.08 and 136.70~\AA\ that they assigned to $3d$--$4f$ transitions, although level assignments were not made. The same lines were identified in low-resolution (around 150 at 130~\AA) solar spectra from the \textit{Extreme Ultraviolet Variability Experiment} \citep[EVE:][]{2012SoPh..275..115W}  by \citet{2011ApJ...740L..52F}, although level information was not provided.

\begin{figure}
    \centering
    \includegraphics[width=0.8\textwidth]{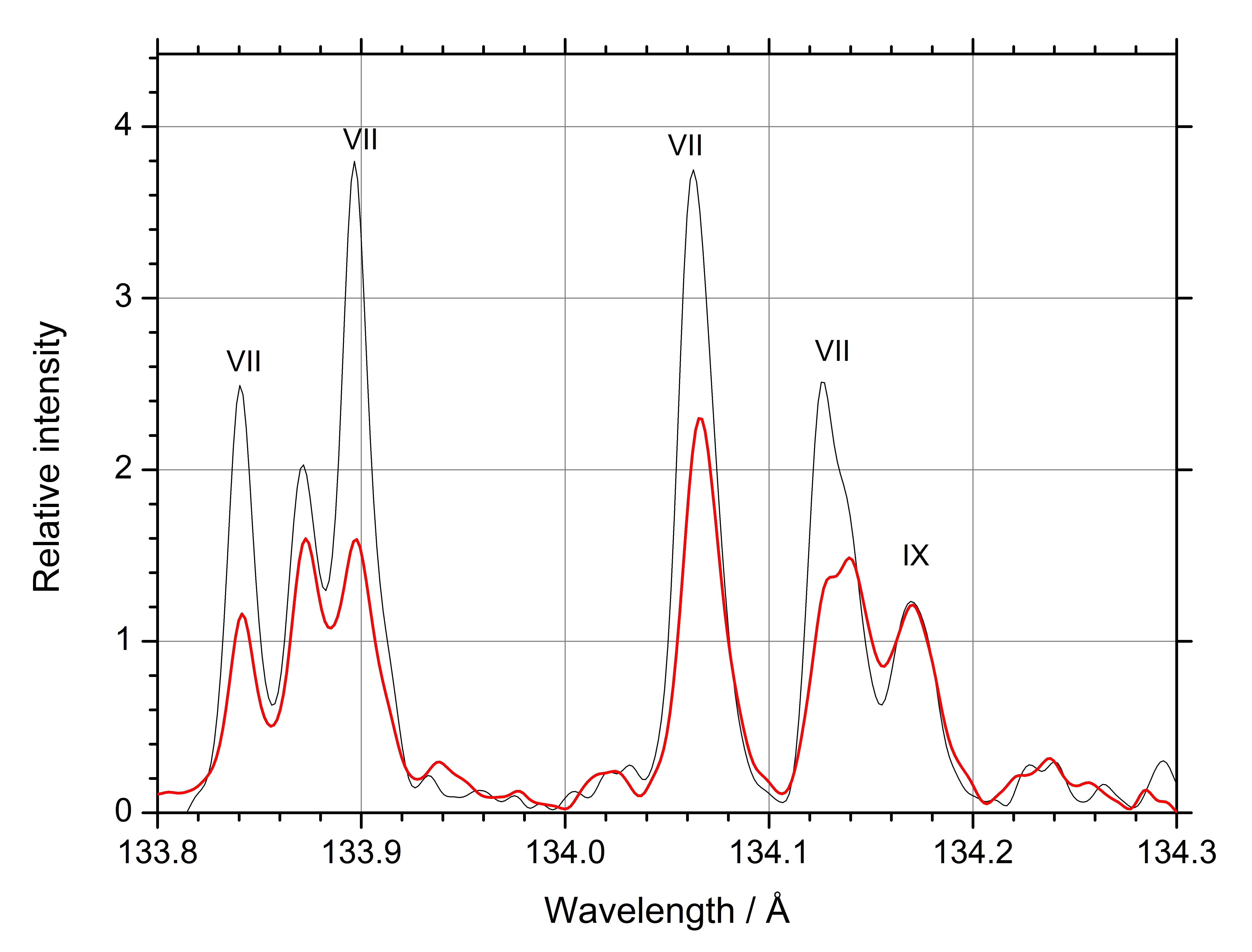}
    \caption{The iron spectrum in the 133.8--134.3~\AA\ range taken in the ``cold" (black) and ``hot" (red) modes of the spark operations (see Figure~\ref{fig.2}).}
    \label{fig.134}
\end{figure}

Our high-resolution spectra reveal several lines near the location of the \citet{2002ApJ...578..648L} 134.08~\AA\ line (Figure~\ref{fig.134}). Of these, the 134.063 and 134.128~\AA\ lines belong to \ion{Fe}{vii} \citep{1981PhyS...23....7E}, while the 134.169~\AA\ line shows behavior consistent with  the \ion{Fe}{ix} lines shown in Figure~\ref{fig.2}.
We identify this line with the $3p^53d$ $^1P_1$ -- $3p^54f$ $^1D_2$ transition, which is supported by the Ritz transition to the $3p^53d$ $^3D_1$ level with a wavelength of 115.042~\AA. The energy of the $3p^54f$ $^1D_2$ level of 1\,329\,872~cm$^{-1}$ is consistent with the energy calculated from the Cowan code. \citet{2012A+A...537A..22O} suggested the $3p^53d$ $^1P_1$ -- $3p^54f$ $^1D_2$  transition belonged to a line at 133.923~\AA. Our spectra have a line at 133.899~\AA\ that is due to \ion{Fe}{vii} \citep{1981PhyS...23....7E}.
Possibly, this \ion{Fe}{vii} lines was identified by \citet{2012A+A...537A..22O} as \ion{Fe}{ix} with a measured value of 133.923~\AA. It should be noted that the other lines at 114.860 and 115.124 Å attributed by \citet{2012A+A...537A..22O} to the transitions from the $3p^54f$ $^1D_2$ level are not present in our spectra.

Our value of 136.674~\AA\ for the $3p^53d$ $^1P_1$ -- $3p^54f$ $^3F_2$ transition is in agreement with the \citet{2002ApJ...578..648L} measurement. The corresponding energy of 1\,316\,205~cm$^{-1}$ for the $3p^54f$ $^3F_2$ level is in good agreement with the value computed from the Cowan code. It is also supported by a Ritz combination to the $3p^53d$ $^3D_1$ level. The corresponding line at 116.881~\AA\ was the only remaining line in this region of the spectrum that had the properties of \ion{Fe}{ix}. On the other hand, \citet{2012A+A...537A..22O} attributed a line at 136.572~\AA\ to the $3p^53d$ $^1P_1$ -- $3p^54f$ $^3F_2$ transition with Ritz support from lines at 113.258 and 116.803~\AA. No \ion{Fe}{ix} lines can  be identified   in our ``hot" spectrum at 136.572 and 113.258~\AA.
The 116.803~\AA\ line, measured in our spectrum at 116.814~\AA, is assigned to the $3p^53d$ $^3D_3$ -- $3p^54f$ $^3F_4$ transition in \ion{Fe}{ix}.

The $3p^54f$ level energies in Table~\ref{tbl.2} were derived from the identified $3p^53d$--$3p^54f$ lines using the program \textsf{LOPT} for least-squares optimization of energy levels \citep{2011CoPhC.182..419K}. The $3p^53d$ levels were fixed to the values of \citet{1978SoPh...57..329E} for the optimization.
The uncertainties of our level energies range from 19 to 40~cm$^{-1}$. Column 2 shows the deviations of the Cowan code energies  from the experimental ones after a fitting of 12 energy levels with four free parameters. The deviations are in good agreement with the estimated level uncertainties. Good agreement is also seen with the energy values found by \citet{1976JOSA...66..240S} and \citet{2012A+A...537A..22O} except for the  questionable cases noted earlier. The differences in the energies are within 60~cm$^{-1}$, which shows that the \citet{1976JOSA...66..240S} uncertainties of  $\pm 200$~cm$^{-1}$ were overestimated.

Table~\ref{tbl.2} shows the designations of the levels in the $LS$ and $JJ$ couplings. It is seen that the $JJ$ coupling representation  is generally better than the $LS$ one. Nevertheless, in Table~\ref{tbl.1} we are using the more convenient $LS$ designations because it is possible to give unambiguously the $LS$ names to all levels although in several cases they are associated with the second component of their wavefunction.

\subsection{The $3p^53d$--$3p^43d^2$ transitions}\label{sect.3d2}

The $3p^43d^2$ comprises 109 fine-structure levels and the $3p^53d$--$3p^43d^2$ transition array gives rise to a large number of emission lines in the 150--270~\AA\ region. \citet{2009ApJ...691L..77Y} was the first to identify lines from this array, using solar spectra obtained with \hinode/EIS. Further identifications from EIS spectra were suggested by \citet{2009ApJ...707..173Y}, \citet{2009A+A...508..501D} and \citet{2014A+A...565A..77D}. Previous laboratory measurements of lines from the $3p^43d^2$ configuration have been made by \citet{2009ApJ...696.2275L} and \citet{2018ApJ...854..114B}.

The wide wavelength coverage and high resolution of our spectra have enabled us to make many new line identifications from the $3p^53d$--$3p^43d^2$ and these are summarized in Table~\ref{tbl.3}. Updated level energies are provided in Table~\ref{tbl.4}.
In total, eighty one lines in the range 151--200 \AA\ were identified, and five of them are doubly classified, i.e., two \ion{Fe}{ix} transitions are assigned to the same observed line (marked with ``db" in column 2 of Table~\ref{tbl.3}). The iron ion spectrum in this region is rich in lines. Depending on excitation conditions in a spark, the lines of \ion{Fe}{vi} to \ion{Fe}{xii} can be present and they can blend, mask or perturb the \ion{Fe}{ix} lines. Even at the high current modes the lines of lower stages of the ionization can be seen due to temporal and spatial inhomogeneity of the spark plasma. The spectra taken at the peak currents 50--100 kA were used for the \ion{Fe}{ix} line measurements. The wavelengths were obtained from the spectra recordings on the photographic plates, while the intensities were taken from the imaging plates, where possible. About 15 \ion{Fe}{ix} lines have  wavelengths close to known \ion{Fe}{vi}--\ion{Fe}{viii} lines \citep{1996PhyS...53..398A,2022ApJS..258...37K,1980OptSp..48..348R} or to the unidentified ``cold" lines that are seen with high intensity in the spectra taken with lower peak currents. It was estimated, after a study of the  behavior of the line intensities of these ion species on the peak current, that in most cases their influence on the \ion{Fe}{ix} lines is negligible. We found and marked in Table~\ref{tbl.3} only six lines that can have contribution to their intensities from ``cold" lines (marked with bl(VII) or bl(VIII) in Table~\ref{tbl.3}) 

The relative intensities are given on a linear scale without accounting for the wavelength dependence of the spectrograph and imaging plate effectivities. The resonance \ion{Fe}{ix} line at 171.073~\AA\ has in our spectrum the intensity 6000 on this scale. The scale is not directly connected to that of the $3p^53d$--$3p^54f$ transitions because the spectra were taken on different recording media and with different excitation in the vacuum spark. According to rough estimates the scales can be different by a factor of two.

Table~\ref{tbl.4} lists the 45 levels in the upper level range of the $3p^43d^2$ configuration (energies above 930\,000~cm$^{-1}$). The 81 lines identified in the present work yield energies for 30 of these levels. Three of the level energies are uncertain (marked with a ? in the table) as they are based on a single questionable line identification. The second column of the table 
compares the observed energies with the energies obtained from Cowan's code using parametric calculations.
The remaining 15 levels in Table~\ref{tbl.4} are listed for completeness and the energies are those from Cowan's code. The values should be valuable for future efforts to identify \ion{Fe}{ix} transitions. Table~\ref{tbl.4} also gives the first three components of each level's eigenvector. The first component of the eigenvector together with the energy level value was used as a label for the transitions in Table~\ref{tbl.3}. 
A difficulty in the level designation should be pointed out. The $3p^43d^2$ configuration consists of two sub-shells with more than one electron. For unambiguous level designation all intermediate quantum numbers should be given: $3p^4(L'S')3d^2(L''S'')LS$. Shorthand designation by the $L'S'$-numbers together with the final $LS$ is adopted in the Cowan code resulting in an ambiguity of the repeated final $LS$-numbers belonging to the same $L'S'$ but to different $L''S''$. The letters $a$ or $b$ are added to distinguish between such cases. The full descriptions for the first components of the eigenvectors are shown in Table~\ref{tbl.4} in square brackets.

The energy parameters after the least-square fitting of the calculated to the experimental energy levels are shown in Table~\ref{tbl.5}. The electrostatic parameters for the configurations with unknown levels as well as the configuration interaction parameters were scaled by a factor 0.85 with respect to the corresponding HFR values \citep[see p.~464 of][]{1981tass.book.....C}.  Only a parameter of interaction between the $3s^23p^43d^2$ and $3s3p^63d$ configurations was fitted for better description of the levels with mixed eigenvectors. The spin-orbit parameters were not scaled, and all of them are omitted from Table~\ref{tbl.5}. The average energies of these configurations were scaled so that their differences with those of the known configurations were approximately the same as in the HFR calculations. The parameters of Table~\ref{tbl.5} were used for the calculations of the energy levels and transition probabilities of the \ion{Fe}{ix} spectrum. The branching ratios for the intensities of the lines from a particular level generally follow the calculated transition probabilities with three exceptions. The intercombination line $^1D_2$--$(^3P)^3Gb$ ($J=3$) at 199.985~\AA\ is too intense and two other intercombination lines $^1D_2$--$(^3P)^3Fa$ ($J=2$) and $^1F_3$--$(^3P)^3Pa$ ($J=2$) at respectively 176.646 and 178.848~\AA\ are too weak in their transition arrays. 


Most of the lines with $gA$ greater than $5\times 10^{10}$~s$^{-1}$  were identified, with the  exceptions of the decays from the $(^3P)^3Da$ term. The three $^3P_{J-1}$--$(^3P)^3D_Ja$ ($J=3,2,1$) transitions have $gA=5.2\times 10^{11}$, $2.9\times 10^{11}$ and $1.4\times 10^{11}$~s$^{-1}$ and calculated  wavelengths 190.07, 189.96 and 190.34~\AA, respectively. The lines may be  blended with lines at 190.037~\AA\ (\ion{Fe}{x}), 189.937~\AA\ (\ion{Fe}{ix}) and 190.296~\AA\ (\ion{Fe}{vi}).

Finally, we give a few remarks about previous measurements and identifications of the \ion{Fe}{ix} lines.

Several laboratory studies of \ion{Fe}{ix} were undertaken with the aid of the Heidelberg \citep{2009ApJ...696.2275L} or the Lawrence Livermore National Laboratory \citep{2012ApJS..201...28B,2018ApJ...854..114B} electron beam ion traps (EBIT). The observed spectra showed the evolution of each ionic stage from Fe$^{5+}$ to at least Fe$^{15+}$ as a function of the electron energy, allowing to distinguish the emission lines from the neighboring ion charge states.

The spectra of \citet{2009ApJ...696.2275L} were recorded in the 125--265~\AA\ range with 0.5--0.8~\AA\ resolution. Four lines (188.5, 189.9, 191.2 and 197.9 \AA) were suggested as belonging to \ion{Fe}{ix}, and these were independently identified by \citet{2009ApJ...691L..77Y} from EIS spectra.  The first three  lines belong to the $3p^53d$ $^3F$ -- $3p^43d^2$ $(^3P)^3G$ multiplet, and the fourth is the $3p^53d$ $^1P_1$ -- $3p^54p$ $^1S_0$ transition.

\citet{2012ApJS..201...28B}  measured with a resolution of about 0.3~\AA\ a dozen spectral features in the 170--200~\AA\ range, some of which could be attributed to \ion{Fe}{ix} based on ionization energy and wavelength coincidences with a \textsf{CHIANTI} spectral model.
Version 7.0 \citep{2012ApJ...744...99L} of  \textsf{CHIANTI} was used, which mostly had only theoretical wavelengths for the lines in this wavelength range and so definitive new identifications could not be made. The authors did provide supporting data for some of the identifications of \citet{2009ApJ...691L..77Y} and \citet{2009ApJ...707..173Y}, however.

\citet{2018ApJ...854..114B} (hereafter BT18) studied emission in the wavelength region 165--175~\AA\ from various species  excited in an EBIT, including \ion{Fe}{ix}. The spectral resolution was 3000 and wavelengths were measured with uncertainties of 10--20~m\AA. The spectrum produced by a beam energy 300~eV yielded lines mainly due to \ion{Fe}{ix} and \ion{Fe}{x}. The authors modeled the \ion{Fe}{ix} emission by employing atomic data computed with the relativistic Multi-Reference M{\o}ller-Plesset (MR-MP) perturbation theory \citep{1999PhRvA..60.2808V} and the Flexible Atomic Code \citep[FAC:][]{2008CaJPh..86..675G}. By comparing their predicted spectra with the EBIT measurements, BT18 were able to match features in their modeled spectra to features in the EBIT spectrum based on proximity in wavelength and intensity. Some of the lines were blended with other species.

The MR-MP method has previously been shown to yield level energies with a spectroscopic accuracy. For example, the $n=3$ levels of \ion{Fe}{xiii} \citep{2004PhRvA..69f2503V} are reproduced to around 0.01\%. However, earlier works did not address complex configurations of the form $3p^k3d^m$ ($k < 6$, $m>1$). Recently, \citet{2020ApJS..247...52S} performed MR-MP calculations for \ion{Fe}{viii} and derived $3p^53d^2$ energies with accuracies up to 0.8\%, which is not high enough for classifying unidentified lines. Although not investigated by BT18, such uncertainties may also apply to their calculations for the $3p^43d^2$ configuration of \ion{Fe}{ix}.

BT18 did not publish their \ion{Fe}{ix} atomic data and so it is not possible to compare radiative decay rates with the present or earlier calculations. However, they do comment that the $3p^43d^2$ $(^1D)^3D_3a$ level (using our level notation) has its strongest decay to $3p^53d$ $^3F_3$ whereas the \textsf{CHIANTI} 8 atomic model has the strongest decay to $3p^53d$ $^3D_3$. The \textsf{CHIANTI} 8 model decay rates are from \citet{2014A+A...565A..77D} and are comparable to the present decay rates. We compared with the calculations of \citet{2002A&A...394..753S} and \citet{2015ApJ...812..174T}, and these both confirmed that the strongest decay is to the $^3D_3$ level. This particular discrepancy affects the identification of the lines at 167.478 and 174.03~\AA\ measured by BT18, as discussed below.

Of the 16 \ion{Fe}{ix} lines identified in Table~5 of BT18, seven  are present in our spectra and we also assign them to \ion{Fe}{ix}. The BT18 line at 172.16~\AA\ was blended with \ion{O}{v} in their spectrum but this is not the case in our spectra. BT18 noted that lines at 170.92 and and 173.90~\AA\ could be due to \ion{Fe}{ix}. Both lines are present in our spectra with \ion{Fe}{ix} properties. All nine of the BT18 measured wavelengths are given in the $\lambda_\mathrm{Pred}$ column of Table~\ref{tbl.3}.

\begin{figure}
    \centering
    \includegraphics[width=0.8\textwidth]{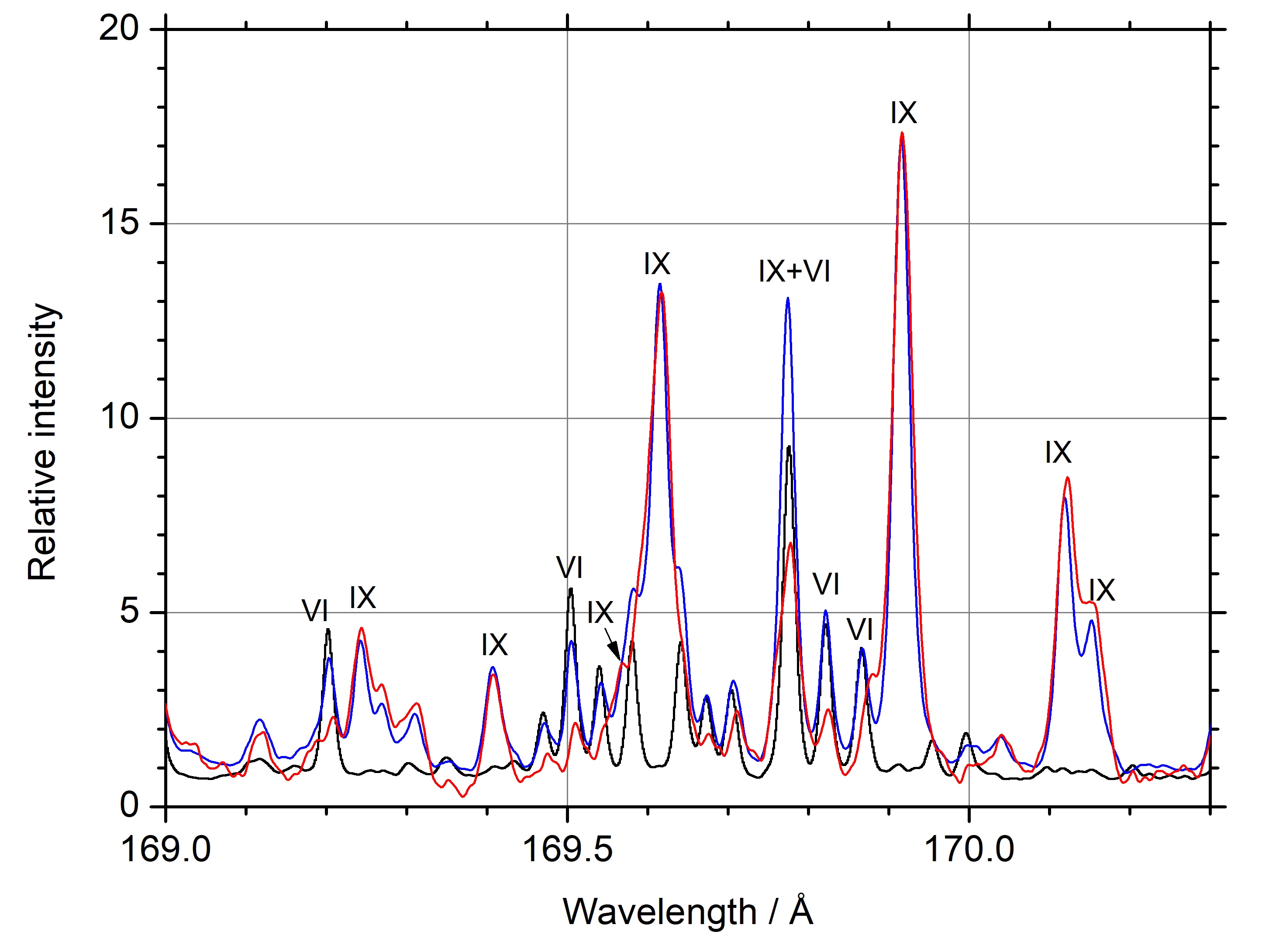}
    \caption{The iron spectrum in the range 169.0--170.3~\AA\ taken at three modes of the spark operation: ``cold" -- black, ``intermediate" -- blue and ``hot" -- red. The lines are marked by the symbols: \textsc{vi} -- \ion{Fe}{vi} (only prominent lines), and \textsc{ix} -- \ion{Fe}{ix}. The relative intensities in arbitrary units are scaled so that the \ion{Fe}{ix} lines  have approximately the same intensities in the ``intermediate" and ``hot" spectra, whereas the \ion{Fe}{vi} lines have approximately the same intensities in the ``cold" and ``intermediate" spectra.
   }
    \label{fig.3}
\end{figure}

The higher resolution of our spectra allow us to resolve the 170.11~\AA\ line of BT18 into two lines at 170.116 and 170.150~\AA\ (see Figure~\ref{fig.3}). The line at 174.03~\AA\ is also resolved into two lines at 174.024 and 174.043~\AA, and the former is a double-blend of two \ion{Fe}{ix} transitions (Table~\ref{tbl.3}).

Of the seven BT18 lines for which they assigned identifications, we confirm only one: the $3p^53d$ $^3D_3$ -- $3p^43d^2$ $(^1D)^3D_3a$ transition at 174.03~\AA. This is the strongest of a pair of transitions that are double-blended in our spectra at 174.024~\AA\ (Table~\ref{tbl.3}). Both identifications are supported by multiple Ritz combinations. Although we agree with the BT18 identification, we note that the BT18 atomic model predicted this transition to be the weakest of all the decays from the $3p^43d^2$ $^3D_3a$ level, yet the measured line is quite strong in their spectra. This is a consequence of the problem with this level noted earlier.

Of the remaining six lines for which identifications disagree, we find two belong to the set of three $3p^53d$ $^3F_J$ -- $3p^43d^2$ $(^3P)^3F_{J}a$ transitions. The BT18 wavelengths 169.605 and 169.900~\AA\ match our wavelengths 169.614 and 169.914~\AA, corresponding to the transitions with $J=3$ and $J=4$, respectively. We identify the $J=2$ transition with a line at 169.773~\AA\ (see discussion below).
Each identification is supported by between four and six Ritz combinations. The 2--3 and 3--4 members of the same multiplet comprise the 170.92~\AA\ feature in the BT18 spectrum, with Ritz wavelengths of 170.918 and 170.927~\AA, respectively. The other Ritz combinations within the 165--175~\AA\ range are at 168.483 and 168.610~\AA\ (Table~\ref{tbl.3}) and lie close to a strong \ion{Fe}{viii} line in the BT18 spectrum and were not reported by BT18.

The identification of the 169.773~\AA\ line warrants further discussion as BT18 stated that this line must come from a spectrum lower than \ion{Fe}{ix}. Figure~\ref{fig.3} shows the spectra in the region of this line taken at different excitation conditions in the spark. The intensities in the spectra are scaled so that their comparison can help in the attribution of the lines to different ions. The 169.773~\AA\ line is clearly blended with a strong \ion{Fe}{vi} line \citep{1996PhyS...53..398A}. But a comparison of the changes of its intensity with those of the other \ion{Fe}{vi} and \ion{Fe}{ix} lines clearly shows that the main contribution to the intensity of this line in ``hot" conditions comes from \ion{Fe}{ix}. The identification  is supported by the observation of an additional five Ritz lines in our spectra. 

Two more of the six BT18 lines for which we have different identifications are 170.11 and 171.685~\AA. These are the two strongest decays of the $(^1D)^3Pa$ ($J=2$) level to $3p^53d$ $^3P_1$ and $^3P_2$, respectively, that we measure at 170.116 and 171.681~\AA. Three additional Ritz combinations outside of the BT18 wavelength range are reported in Table~\ref{tbl.3}. The remaining two BT18 lines are at 171.26 and 172.16~\AA\ and we identify them with decays from the $(^3P)^1F_3$ level. Our wavelengths are 171.279~\AA\ and 172.219~\AA\ and they correspond to decays to the $3p^53d$ $^3D_2$ and $^1F_3$ levels, respectively. Note the latter is blended with \ion{O}{v} in the BT18 spectrum, hence the wavelength discrepancy. A third decay to the $^1D_2$ level can be identified as an enhanced wing to the strong 169.614~\AA\ line, as indicated in Figure~\ref{fig.3}.

BT18 listed two strong transitions at 167.478 and 167.654~\AA\ as blends of \ion{Fe}{ix} with \ion{Fe}{viii}. The former identification was made on account of the BT18 atomic model predicting the strongest transition from the $3p^43d^2$ $(^1D)^3D_3a$ level near this wavelength. As noted earlier this prediction is at odds with other atomic calculations, and our spectra do not suggest a contribution from \ion{Fe}{ix}. A similar problem may affect the BT18 identification of the 167.654~\AA\ line. That is, if their atomic model predicts the $3p^53d$ $^3F_2$ -- $3p^43d^2$ $^3D_2a$ to be the strongest decay, then this disagrees with our atomic calculations and those in \textsf{CHIANTI}. As with the 167.478~\AA\ line, we do not find evidence of an \ion{Fe}{ix} contribution.

The remaining seven \ion{Fe}{ix} lines listed by BT18 that are not found in our spectra were weak in their spectrum (intensities $\le 0.1$ on the BT18 scale). They could belong to other stages of ionization, but the different excitation conditions in the EBIT compared to our spark spectrum could also be responsible for the differences. The EBIT plasma has a lower density that is more typical of the solar corona.

A breakthrough in the analyses of the \ion{Fe}{ix} $3p^53d$ - $3p^43d^2$ lines in the Sun's spectrum came from observations by \textit{Hinode}/EIS. The two wavelength bands 170--212 and 246--292~\AA\ are observed with a spectral resolution of around 0.06~\AA. Imaging capability is critical to discriminating between neighboring ionization stages, as demonstrated by \citet{2009ApJ...691L..77Y} who compared images of coronal loop structures in lines of \ion{Fe}{viii} through \ion{Fe}{x}.

\citet{2009ApJ...706....1L}  created an atlas of the Sun’s spectral lines of ions formed between $10^5$~K and $10^6$~K from a bright point related to the footpoint region of a coronal loop. The intensities of ``cold" lines were enhanced over normal values in this atlas. It permitted to make the first identification of three lines from the $3p^53d$--$3p^43d^2$ transitions, namely, the main lines from the $3p^53d$ $^3F$ -- $3p^43d^2$ $(^3P)^3G$ multiplet \citep{2009ApJ...691L..77Y} mentioned above. In an extension of this result, \citet{2009ApJ...707..173Y} found the Ritz combinations for the $3p^43d^2$ $(^3P)^3G_{4,3}$ levels, thus confirming their identifications. Three lines (176.959, 177.594 and 199.986 \AA) were added as the transitions to the $3p^53d$ $^3F_{3,4}$ levels from the $3p^43d^2$ $(^1D)^3D_{3,2}$ levels. It should be noted that the first line in the EIS spectrum is not fully resolved from the \ion{Fe}{vii} 176.905 \AA\ line. The wavelength of this line is 176.978~\AA\ from our fully resolved spectrum. Two other lines at 178.699 and 178.985 A were tentatively suggested as the $3p^53d$ $^3F_2$--$3p^43d^2$ $(^1D)^3D_1$ and $3p^53d$ $^3D_3$--$3p^43d^2$ $(^3P)^3F$a ($J=4$) transitions, respectively. We have confirmed these suggestions. In summary, ten lines in the $3p^53d$--$3p^43d^2$ array of \ion{Fe}{ix} were identified with specific $3p^43d^2$ levels.

\citet{2009ApJ...707..173Y} published a list of seven additional observed lines that were suggested as being due to \ion{Fe}{ix} based on image morphology, but for which transition identifications could not be assigned.
We have identified all but one of these lines and marked  them with YLTW  in Table~\ref{tbl.3}. On a basis of our laboratory spectrum, we  have identified three additional lines in the \citet{2009ApJ...706....1L}  atlas. The 182.158, 188.823 and 194.806 \AA\ EIS lines are identified respectively with  $3p^33d$ $^3F_4$ -- $3p^43d^2$ $(^3P)^3Fb$ ($J=4$), $3p^33d$ $^1F_3$ -- $3p^43d^2$ $(^1S)^1G_4$ and $3p^33d$ $^3D_2$ -- $3p^43d^2$ $(^3P)^3Fb$ ($J=3$) transitions. The latter  is not fully resolved from the \ion{Fe}{vii} 194.770 \AA\ line in the EIS spectrum, and our laboratory wavelength is 194.796~\AA. Note that the wavelengths of all the EIS \ion{Fe}{ix} lines from the \citet{2009ApJ...706....1L} atlas listed in Table~\ref{tbl.3} are corrected for the red shift with velocity 16~km~s$^{-1}$ as suggested by \citet{2009ApJ...707..173Y}.

\citet{2009A+A...508..501D} published a spectral atlas from a coronal loop rooted in a sunspot that also exhibited strongly enhanced cool emission lines.
He confirmed the \citet{2009ApJ...691L..77Y} identifications of the three main lines of the $3p^53d$ $^3F$ -- $3p^43d^2$ $(^3P)^3G$ multiplet and he tentatively identified three weak transitions from the $3p^43d^2$ configuration levels that we reject. However, the line list contains many other lines left unidentified or considered as blends that are also present in the \citet{2009ApJ...706....1L} atlas and are identified in our laboratory spectrum. These lines are marked with D09 in Table~\ref{tbl.3}. Their wavelengths were corrected for a redshift of  10~km~s$^{-1}$.

Large scale intermediate-coupling $R$-matrix scattering calculations for electron collisional excitation of \ion{Fe}{ix} were performed by \citet{2014A+A...565A..77D}. The data were used to create a new atomic model for the ion, and intensities were computed and compared with observations.
Good agreement with the known \ion{Fe}{ix} lines in the EIS spectrum was obtained. Based on this agreement a few new weak \ion{Fe}{ix} lines were tentatively identified  \citep{2009A+A...508..501D}. We confirmed the identification of one line at 192.630~\AA\ as the $3p^33d$ $^3D_3$--$3p^43d^2$ $(^3P)^3Fb$ ($J=4$) transition. The identification of the 194.784~\AA\ line was changed. The other suggested  lines are absent in our laboratory spectra, perhaps because of different excitation conditions in the Sun and our spark spectra or they belong to \ion{Fe}{vii} \citep{1981PhyS...23....7E,2022ApJS..258...37K}.

The $3p^53d$--$3p^54p$ lines should be also located in the studied wavelength range. As mentioned above, a line at 197.862~\AA\ in the EIS spectrum was identified as the $3p^53d$ $^1P_1$ -- $3p^54p$ $^1S_0$ transition by \citet{2009ApJ...691L..77Y}. This identification was supported by the observation of two lines at 717.661 and 803.422~\AA, corresponding to decays  to the $3p^54s$ $^1P_1$ and $^3P_1$ levels \citep{2009ApJ...707.1191L}. These lines were measured with the \textit{Solar Ultraviolet Measurements of Emitted Radiation} \citep[SUMER:][]{1995SoPh..162..189W} instrument. The 197.862~\AA\ line is very weak in our spectrum. We did not succeed in finding any other lines of the $3p^53d$ - $3p^54p$ transitions, which possess smaller transition probabilities than the 197.862~\AA\ line. Since only one upper level is known in the $3p^54p$ configuration, we considered this configuration as ``unknown" in our calculations, scaling only its average energy by a predetermined factor (Table~\ref{tbl.5}). In this approach, the calculated energy of the $3p^54p$ $^1S_0$ level is lower than the experimental one by about 3900~cm$^{-1}$.

\section{Conclusions}\label{sect.conc}

Using high-resolution laboratory spectra in the 110--200 \AA\ range the analysis of energy levels and spectral lines of \ion{Fe}{ix} was greatly extended. Many weak lines were added to the previous analyses of the $3p^53d$--$3p^54f$ transitions extending to 25 the number of identified lines. Seventy-three lines of the $3p^53d$--$3p^43d^2$ transition array were identified, bringing to 81 the number of known lines in this transition array. A number of lines assigned to \ion{Fe}{ix} were identified in \textit{Hinode}/EIS spectra. The data can be used for diagnostics of solar plasma and provide a benchmark for further development of atomic theory.

\begin{acknowledgements}
P.R.~Young acknowledges  support from the NASA Heliophysics Data Environment Enhancements program and the NASA Individual Scientist Funding Model competitive work package program. \textsf{CHIANTI} is a collaborative project involving NASA Goddard Space Flight Center, George Mason University, the University of Michigan (USA), and the University of Cambridge (UK).
\end{acknowledgements}

\bibliography{ms}{}
\bibliographystyle{aasjournal}

\movetabledown=1.5cm
\begin{rotatetable}
\begin{deluxetable}{ccccccccccCccCc}
\tabletypesize{\footnotesize}
\tablecaption{Wavelengths of the $3p^53d$--$3p^54f$ transitions in \ion{Fe}{ix}.\label{tbl.1}}
\tablehead{ & $gA$\tablenotemark{b} & \multicolumn{7}{c}{$\lambda$\tablenotemark{c} (\AA)} & &
  \multicolumn{2}{c}{$3p^53d$} & & \multicolumn{2}{c}{$3p^54f$} \\
\cline{3-9} \cline{11-12} \cline{14-15}
  $I$\tablenotemark{a} & ($10^{10}$~s$^{-1}$) & This work & Ritz\tablenotemark{d} & TW-Ritz\tablenotemark{e} & S76 & TW-S76 & O'D12 & TW-O'D12 && \mathrm{Term} & $E$ (cm$^{-1}$) && \mathrm{Term} & $E$ (cm$^{-1}$) }
\startdata
21 & 22 & 111.664 & 111.665(3) & -0.001 &         &        &
&        &  &  ^3F_3 & 429310.9 &  & ^1G_4 & 1324844 \\
27 & 35 & 111.692 & 111.692(4) & 0.000 & 111.689 & 0.003 &  111.713 & -0.021 &  &  ^3P_0 & 405772 &  & ^3D_1 & 1301088\\
51 & 73 & 111.785 & 111.790(3) & -0.005 & 111.795 & -0.01 & 111.791 & -0.006 &  &  ^3P_1 & 408315.1 &  & ^3D_2 & 1302850\\
24 & 29 & 112.011 VII  & 112.011(4) & 0.000 & 112.017 & -0.006 &  112.031 & -0.020 &  &  ^3P_1 & 408315.1 &  & ^3D_1 & 1301088\\
108 & 126 & 112.093 & 112.096(3) & -0.003 & 112.097 & -0.004 & 112.096 & -0.003 &  &  ^3P_2 & 413669.2 &  & ^3D_3 & 1305762\\
29 & 38 & 112.375 & 112.380(3) & -0.005 & 112.375 & 0 &           &        &  &  ^3F_2 & 433818.8 &  & ^3F_3 & 1323654\\
  - & 30 &    -    VIII & 112.463(3) &        &         &        & 112.464 &        &  &  ^3P_2 & 413669.2 &  & ^3D_2 & 1302850\\
16 & 18 & 112.875 & 112.880(3) & -0.005 &         &        &           &        &  &  ^3F_4 & 425809.8 &  & ^3F_4 & 1311707\\
20 & 24 & 113.529 & 113.529(4) & 0.000 &         &        &           &        &  &  ^3F_3 & 429310.9 &  & ^3G_3 & 1310147\\
18 & 15 & 113.57 & 113.572(3) & -0.002 &         &        & 113.571 & -0.001 &  &  ^3F_4 & 425809.8 &  & ^3G_4 & 1306305\\
250 & 266 & 113.789 & 113.789(5) &        & 113.792 & -0.003 & 113.793 & -0.004 &  &  ^3F_4 & 425809.8 &  & ^3G_5 & 1304630\\
136 & 172 & 114.024 & 114.026(3) & -0.002 & 114.027 & -0.003 & 114.024 & 0 &  &  ^3F_3 & 429310.9 &  & ^3G_4 & 1306305\\
98 & 113 & 114.112 & 114.113(4) & -0.001 & 114.116 & -0.004 & 114.111 & 0.001 &  &  ^3F_2 & 433818.8 &  & ^3G_3 & 1310147\\
30 & 12 & 115.042 VII  & 115.044(3) & -0.002 &         &        &           &        &  &  ^3D_3 & 455612.2 &  & ^1G_4 & 1324844\\
30 & 34 & 115.042 VII  & 115.041(3) & 0.001 &         &        &           &        &  &  ^3D_1 & 460616 &  & ^1D_2 & 1329872\\
51 & 112 & 115.354 & 115.353(3) & 0.001 & 115.344 & 0.01 & 115.353 & 0.001 &  &  ^1D_2 & 456752.7 &  & ^3F_3 & 1323654\\
84 & 130 & 116.002 & 116.002(5) &        & 116.001 & 0.001 & 115.996 & 0.006 &  &  ^3D_2 & 462616.6 &  & ^1F_3 & 1324670\\
101 & 172 & 116.414 & 116.412(3) & 0.000 & 116.408 & 0.006 & 116.408 & 0.006 &  &  ^1F_3 & 465828.4 &  & ^1G_4 & 1324844\\
10 & 9 & 116.578 & 116.574(3) & 0.004 &         &        &           &        &  &  ^1F_3 & 465828.4 &  & ^3F_3 & 1323654\\
171 & 171 & 116.814 VII  & 116.809(4) & 0.005 & 116.806 & 0.008 & 116.803 & 0.011 &  &  ^3D_3 & 455612.2 &  & ^3F_4 & 1311707\\
34 & 57 & 116.881 VII  & 116.879(3) & 0.002 &         &        &           &        &  &  ^3D_1 & 460616 &  & ^3F_2 & 1316205\\
25 & 32 & 117.629 & 117.626(4) & 0.003 &         &        & 117.626 & 0.003 &  &  ^3D_3 & 455612.2 &  & ^3D_3 & 1305762\\
19 & 21 & 118.984 & 118.980(3) & 0.004 &         &        & 118.978 & 0.006 &  &  ^1F_3 & 465828.4 &  & ^3G_4 & 1306305\\
10 & 11 & 119.019 & 119.015(4) & 0.004 &         &        &           &        &  &  ^3D_2 & 462616.6 &  & ^3D_2 & 1302850\\
35 & 72 & 134.169 & 134.169(4) & 0.000 & 134.743? & -0.574       & 133.923  &  0.246      &  &  ^1P_1 & 584546 &  & ^1D_2 & 1329872\\
21 & 44 & 136.674 VII  & 136.676(4) & -0.002 &         &        &  136.572  &  0.102       &  &  ^1P_1 & 584546 &  & ^3F_2 & 1316205\\
\enddata
\tablenotetext{a}{Observed relative intensity in arbitrary units (see text).}
\tablenotetext{b}{Product of upper level weight $g$ and transition probability $A$ calculated with eigenfunctions obtained from the fitting of the calculated to experimental level energies.}
\tablenotetext{c}{Symbols: VII -- line is present in a list of the \ion{Fe}{vii} lines reported by \citet{1981PhyS...23....7E}, an influence on the intensity and the wavelength of the corresponding \ion{Fe}{ix} line is negligible at present excitation conditions; VIII – line is masked by a wing of the \ion{Fe}{viii} line at 112.472~\AA. Previous measurements: S76 -- \citet{1976JOSA...66..240S}; O'D12 -- \citet{2012A+A...537A..22O}.}
\tablenotetext{d}{Wavelength derived from the final level energies (Ritz wavelength).}
\tablenotetext{e}{A blank value for the TW--Ritz difference indicates that the upper level is derived from that line only.}
\end{deluxetable}
\end{rotatetable}

\pagebreak

\begin{deluxetable*}{cccccccCCcc}
\tabletypesize{\small}
\tablecaption{Energy levels of the $3p^54f$ configuration of \ion{Fe}{ix}\label{tbl.2}}
\tablehead{
  \multicolumn6c{$E$\tablenotemark{a} (cm$^{-1}$)} &&
  \multicolumn{2}{c}{Designation\tablenotemark{c}}\\
  \cline{1-6} \cline{8-10}
  \colhead{TW} & o-c\tablenotemark{b} & S76 & TW-S76 & O'D12 & TW-O'D12 &
  & LS  & JJ && \colhead{N\tablenotemark{d}}
}
\startdata
1301088(23) & 24 & 1301070 & 18 & 1300923? & 165 &   & 99\%\ ^3D_1 & 99\%\ (3/2, 5/2)_1 &   & 2 \\
1302850(30) & -20 & 1302800 & 50 & 1302841 & 9 &   & 93\%\ ^3D_2 & 53\%\ (3/2, 5/2)_2 &   & 2 \\
1304630(30) & 50 & 1304595 & 35 & 1304598 & 32 &   & 99\%\ ^3G_5 & 99\%\ (3/2, 7/2)_5 &   & 1 \\
1305762(30) & -38 & 1305750 & 12 & 1305762 & 0 &   & 79\%\ ^3D_3 & 79\%\ (3/2, 7/2)_3 &   & 2 \\
1306305(21) & -12 & 1306295 & 10 & 1306319 & -14 &   & 69\%\ ^3G_4 & 96\%\ (3/2, 5/2)_4 &   & 3 \\
1310147(22) & 10 & 1310110 & 37 & 1310158 & -11 &   & 56\%\ ^3G_3 & 81\%\ (3/2, 5/2)_3 &   & 2 \\
1311707(30) & -24 & 1311730 & -23 & 1311755 & -48 &   & 50\%\ ^3F_4 & 95\%\ (3/2, 7/2)_4 &   & 2 \\
1316205(19) & 15 &    \nodata    &  \nodata   & 1316758 & -553 &   & 58\%\ ^3F_2 & 49\%\ (3/2, 7/2)_2 &   & 2 \\
1323654(24) & 17 & 1323715 & -61 & 1323657 & -3 &   & 28\%\ ^3F_3 & 86\%\ (1/2, 5/2)_3 &   & 3 \\
1324670(30) & -42 & 1324680 & -10 & 1324715 & -45 &   & 38\%\ ^1F_3 & 83\%\ (1/2, 7/2)_3 &   & 1 \\
1324844(22) & 46 & 1324885 & -41 & 1324876 & -32 &   & 30\%\ ^1G_4 & 93\%\ (1/2, 7/2)_4 &   & 3 \\
1329872(19) & -20 & 1326700? & 3172 & 1331244 & -1372 &   & 55\%\ ^1D_2
& 89\%\ (1/2, 5/2)_2 &   & 2 \\
\enddata
\tablenotetext{a}{Key to references for the energy level values: O'D12 -- \citet{2012A+A...537A..22O}, S76 -- \citet{1976JOSA...66..240S}, TW -- this work.}
\tablenotetext{b}{Residual (``obs. - calc.") of the parametric least-squares fit with Cowan's codes.}
\tablenotetext{c}{Percentages of the leading components in the eigenvector of each level in $LS$- and $JJ$-coupling. In $LS$-coupling, the leading components of the 1\,323\,654, 1\,324\,670 and 1\,324\,844~cm$^{-1}$ levels correspond to the second term of the eigenvector composition.}
\tablenotetext{d}{Number of observed lines determining the level value in the least-squares optimization procedure.}
\end{deluxetable*}

\pagebreak

\startlongtable
\begin{deluxetable}{ccccccccccccccc}
\tabletypesize{\footnotesize}
\tablecaption{Wavelengths of the $3p^53d$--$3p^43d^2$ transitions in \ion{Fe}{ix}. \label{tbl.3}}
\tablehead{
   &&&&&&\multicolumn3c{$3p^53d$} && 
   \multicolumn3c{$3p^43d^2$} \\
   \cline{7-9} \cline{11-13}
   \colhead{$\lambda$ (\AA)} &
   \colhead{Char.\tablenotemark{a}} &
   \colhead{$I$\tablenotemark{b}} &
   \colhead{$gA$\tablenotemark{c}} & 
   \colhead{$\lambda_\mathrm{Ritz}$ (\AA)\tablenotemark{d}} &
   \colhead{o--c\tablenotemark{e}} &
   \colhead{Term} &
   \colhead{$J$} &
   \colhead{$E$ (cm$^{-1}$)} & &
   \colhead{Term\tablenotemark{f}} &
   \colhead{$J$} &
   \colhead{$E$ (cm$^{-1}$)} &
   \colhead{$\lambda_\mathrm{Prev}$ (\AA)\tablenotemark{g}}
}
\startdata
151.549 &          & 200 & 205 & 151.550(2) & -0.001 & $^1D$ & 2 & 456752.7 &  & $(^3P)^1D a$ & 2 & 1116602 &              &             \\
152.910 &          & 80 & 73 & 152.909(2) & 0.001 & $^3D$ & 2 & 462616.6 &  & $(^3P)^1D a$ & 2 & 1116602 &              &             \\
158.756 &          & 50 & 48 & 158.759(3) & -0.003 & $^3P$ & 0 & 405772 &  & $(^1D)^3D a$ & 1 & 1035659 &              &             \\
159.011 &          & 130 & 101 & 159.014(2) & -0.003 & $^3P$ & 1 & 408315.1 &  & $(^1D)^3D a$ & 2 & 1037191 &              &             \\
160.377 &          & 80 & 39 & 160.379(2) & -0.002 & $^3P$ & 2 & 413669.2 &  & $(^1D)^3D a$ & 2 & 1037191 &              &             \\
162.178 &          & 400 & 300 & 162.183(3) & -0.005 & $^3P$ & 2 & 413669.2 &  & $(^1D)^3D a$ & 3 & 1030255 &              &             \\
162.720 &          & 60 & 33 & 162.723(2) & -0.003 & $^3P$ & 1 & 408315.1 &  & $(^3P)^3F a$ & 2 & 1022856 &              &             \\
165.439 & ?        & 130 & 790 & 165.439(5) &        & $^1P$ & 1 & 584546 &  & $(^3P)^1S$  & 0 & 1188998 &              &             \\
168.483 &          & 70 & 91 & 168.479(2) & 0.004 & $^3F$ & 3 & 429310.9 &  & $(^3P)^3F a$ & 2 & 1022856 &              &             \\
168.610 &          & 140 & 88 & 168.610(2) & 0.000 & $^3F$ & 4 & 425809.8 &  & $(^3P)^3F a$ & 3 & 1018894 &              &             \\
169.242 &          & 260 & 138 & 169.245(2) & -0.003 & $^3D$ & 3 & 455612.2 &  & $(^3P)^1F$  & 3 & 1046472 &              &             \\
169.407 &          & 160 & 201 & 169.410(3) & -0.003 & $^3P$ & 0 & 405772 &  & $(^1D)^3P a$ & 1 & 996057 &              &             \\
169.579 & p        & 420 & 259 & 169.572(2) & 0.007 & $^1D$ & 2 & 456752.7 &  & $(^3P)^1F$  & 3 & 1046472 &              &             \\
169.614 &          & 1240 & 1263 & 169.611(2) & 0.003 & $^3F$ & 3 & 429310.9 &  & $(^3P)^3F a$ & 3 & 1018894 & 169.605 BT18TW &             \\
169.773 & bl(VII)      & $\sim$500 & 744 & 169.769(2) & 0.004 & $^3F$ & 2 & 433818.8 &  & $(^3P)^3F a$ & 2 & 1022856 &              &             \\
169.914 &          & 1380 & 1776 & 169.910(2) & 0.004 & $^3F$ & 4 & 425809.8 &  & $(^3P)^3F a$ & 4 & 1014356 & 169.900 BT18TW &             \\
170.116 &          & 600 & 383 & 170.117(3) & -0.001 & $^3P$ & 1 & 408315.1 &  & $(^1D)^3P a$ & 2 & 996147 & 170.11  BT18TW &             \\
170.150 & w        & 290 & 87 & 170.143(3) & 0.007 & $^3P$ & 1 & 408315.1 &  & $(^1D)^3P a$ & 1 & 996057 &              &             \\
170.889 & ?        & 550 & 292 & 170.889(4) &        & $^3P$ & 1 & 408315.1 &  & $(^1D)^3P a$ & 0 & 993490 &              &             \\
170.924 & db       & 740 & 158 & 170.918(2) & 0.006 & $^3F$ & 2 & 433818.8 &  & $(^3P)^3F a$ & 3 & 1018894 & 170.92  BT18TW &             \\
170.924 & db       & 740 & 197 & 170.927(2) & -0.003 & $^3F$ & 3 & 429310.9 &  & $(^3P)^3F a$ & 4 & 1014356 & 170.92  BT18TW &             \\
171.279 &          & 450 & 137 & 171.275(2) & 0.004 & $^3D$ & 2 & 462616.6 &  & $(^3P)^1F$  & 3 & 1046472 & 171.26  BT18TW           &             \\
171.681 & bl(VII)  & 1030 & 1048 & 171.680(3) & 0.001 & $^3P$ & 2 & 413669.2 &  & $(^1D)^3P a$ & 2 & 996147 & 171.685 BT18TW &             \\
171.706 & bl(VII)  & 940 & 522 & 171.707(3) & -0.001 & $^3P$ & 2 & 413669.2 &  & $(^1D)^3P a$ & 1 & 996057 &              &             \\
171.948 &          & 170 & 111 & 171.946(2) & 0.002 & $^3D$ & 3 & 455612.2 &  & $(^1D)^3D a$ & 2 & 1037191 &              &             \\
172.219 &          & 920 & 1180 & 172.223(2) & -0.004 & $^1F$ & 3 & 465828.4 &  & $(^3P)^1F$  & 3 & 1046472 & 172.16\tablenotemark{i} BT18TW &             \\
172.285 &          & 100 & 96 & 172.284(2) & 0.001 & $^1D$ & 2 & 456752.7 &  & $(^1D)^3D a$ & 2 & 1037191 &              &             \\
172.482 &          & 390 & 44 & 172.486(3) & -0.004 & $^3P$ & 2 & 413669.2 &  & $(^1D)^3Db$ & 1 & 993425 &              &             \\
172.795 &          & 70 & 33 & 172.796(3) & -0.001 & $^3P$ & 2 & 413669.2 &  & $(^1D)^3Db$ & 2 & 992387 &              &             \\
173.258 &          & 200 & 180 & 173.255(3) & 0.003 & $^3P$ & 0 & 405772 &  & $(^3P)^3S$  & 1 & 982956 &              &             \\
173.902 &          & 780 & 537 & 173.900(3) & 0.002 & $^3D$ & 1 & 460616 &  & $(^1D)^3D a$ & 1 & 1035659 & 173.902 BT18TW &             \\
174.024 & db       & 1530 & 1578 & 174.020(3) & 0.004 & $^3D$ & 3 & 455612.2 &  & $(^1D)^3D a$ & 3 & 1030255 & 174.03  BT18TW &             \\
174.024 & db       & 1530 & 359 & 174.022(4) & 0.002 & $^3P$ & 1 & 408315.1 &  & $(^3P)^3S$  & 1 & 982956 & 174.03  BT18TW &             \\
174.043 &          & 1400 & 945 & 174.042(2) & 0.001 & $^3D$ & 2 & 462616.6 &  & $(^1D)^3D a$ & 2 & 1037191 & 174 03  BT18TW &             \\
175.020 & db       & 530 & 121 & 175.022(3) & -0.002 & $^1D$ & 2 & 456752.7 &  & $(^1D)^1D a$ & 2 & 1028110 &              &             \\
175.020 & db       & 530 & 770 & 175.020(2) & 0.000 & $^1F$ & 3 & 465828.4 &  & $(^1D)^3D a$ & 2 & 1037191 &              &             \\
175.551 &          & 60 & 26 & 175.552(2) & -0.001 & $^3F$ & 4 & 425809.8 &  & $(^1S)^1G$  & 4 & 995441 &              &             \\
175.648 & db       & 420 & 409 & 175.642(3) & 0.006 & $^3D$ & 3 & 455612.2 &  & $(^3P)^3P a$ & 2 & 1024953 &              &             \\
175.648 & db       & 420 & 186 & 175.658(4) & -0.010 & $^3P$ & 2 & 413669.2 &  & $(^3P)^3S$  & 1 & 982956 &              &             \\
175.712 &          & 300 & 201 & 175.718(3) & -0.006 & $^3D$ & 1 & 460616 &  & $(^3P)^3P a$ & 1 & 1029709 &              &             \\
175.774 & ?        & 240 & 155 & 175.774(5) &        & $^3D$ & 1 & 460616 &  & $(^1S)^1S$  & 0 & 1029528 &              &             \\
175.993 &          & 140 & 147 & 175.994(3) & -0.001 & $^1D$ & 2 & 456752.7 &  & $(^3P)^3P a$ & 2 & 1024953 &              &             \\
176.289 &          & 320 & 77 & 176.291(2) & -0.002 & $^3D$ & 3 & 455612.2 &  & $(^3P)^3F a$ & 2 & 1022856 &              &             \\
176.344 & bl(VII)  & 510 & 433 & 176.338(3) & 0.006 & $^3D$ & 2 & 462616.6 &  & $(^3P)^3P a$ & 1 & 1029709 &              &             \\
176.599 & ?bl(VII) & 520 & 632 & 176.599(5) &        & $^1D$ & 2 & 456752.7 &  & $(^3P)^1P$  & 1 & 1023007 &              &             \\
176.646 &          & 70 & 148 & 176.646(2) & 0.000 & $^1D$ & 2 & 456752.7 &  & $(^3P)^3F a$ & 2 & 1022856 &              &             \\
176.840 &          & 240 & 124 & 176.837(3) & 0.003 & $^3D$ & 2 & 462616.6 &  & $(^1D)^1D a$ & 2 & 1028110 &              &             \\
176.978 &          & 1210 & 1598 & 176.978(5) &        & $^3F$ & 4 & 425809.8 &  & $(^1D)^3Db$ & 3 & 990852 & 176.959\tablenotemark{h} YL  & 176.945\tablenotemark{h} D14 \\
177.173 &  bl(VII)     & 850 & 179 & 177.171(3) & 0.002 & $^1F$ & 3 & 465828.4 &  & $(^1D)^3D a$ & 3 &  1030255 &              &             \\
177.599 &          & 780 & 958 & 177.596(3) & 0.003 & $^3F$ & 3 & 429310.9 &  & $(^1D)^3Db$ & 2 & 992387 & 177.594 YL   & 177.592 D14 \\
177.845 &          & 300 & 234 & 177.847(3) & -0.002 & $^1F$ & 3 & 465828.4 &  & $(^1D)^1D a$ & 2 & 1028110 &              &             \\
177.856 & db       & 300 & 325 & 177.860(2) & -0.004 & $^3D$ & 1 & 460616 &  & $(^3P)^3F a$ & 2 & 1022856 &              &             \\
177.856 & db       & 300 & 47 & 177.861(3) & -0.005 & $^3F$ & 2 & 433818.8 &  & $(^1D)^3P a$ & 1 & 996057 &              &             \\
177.888 &          & 270 & 151 & 177.891(2) & -0.003 & $^1D$ & 2 & 456752.7 &  & $(^3P)^3F a$ & 3 & 1018894 &              &             \\
178.059 & bl(XI)   & 760 & 606 & 178.065(3) & -0.006 & $^3D$ & 3 & 455612.2 &  & $(^3P)^1Db$ & 2 & 1017204 &              &             \\
178.431 &          & 40 & 60 & 178.428(3) & 0.003 & $^1D$ & 2 & 456752.7 &  & $(^3P)^1Db$ & 2 & 1017204 &              &             \\
178.701 &          & 930 & 626 & 178.697(3) & 0.004 & $^3F$ & 2 & 433818.8 &  & $(^1D)^3Db$ & 1 & 993425 & 178.699 YL   &             \\
178.848 &          & 80 & 379 & 178.851(3) & -0.003 & $^1F$ & 3 & 465828.4 &  & $(^3P)^3P a$ & 2 & 1024953 &              &             \\
178.972 &          & 810 & 492 & 178.973(2) & -0.001 & $^3D$ & 3 & 455612.2 &  & $(^3P)^3F a$ & 4 & 1014356 & 178.985 YL   &             \\
179.027 &          & 530 & 69 & 179.029(3) & -0.002 & $^3F$ & 2 & 433818.8 &  & $(^1D)^3Db$ & 2 & 992387 &              &             \\
179.753 & bl(XI)   & 1110 & 319 & 179.767(3) & -0.014 & $^3D$ & 2 & 462616.6 &  & $(^3P)^3F a$ & 3 & 1018894 &              &             \\
181.367 &          & 320 & 460 & 181.364(3) & 0.003 & $^1F$ & 3 & 465828.4 &  & $(^3P)^1Db$ & 2 & 1017204 &              &             \\
182.161 & bl(XI)   & 1370 & 350 & 182.172(3) & -0.011 & $^3F$ & 4 & 425809.8 &  & $(^3P)^3Fb$ & 4 & 974743 & 182.158 LYTW & 182.146 D09 \\
182.305 & bl       & 870 & 161 & 182.306(3) & -0.001 & $^1F$ & 3 & 465828.4 &  & $(^3P)^3F a$ & 4 & 1014356 &              &             \\
182.919 &          & 760 & 217 & 182.928(2) & -0.009 & $^3F$ & 3 & 429310.9 &  & $(^3P)^3Fb$ & 3 & 975974 &              &             \\
184.442 &          & 300 & 56 & 184.449(2) & -0.007 & $^3F$ & 2 & 433818.8 &  & $(^3P)^3Fb$ & 3 &  975974&              &             \\
185.244 & p(VIII)  & 650 & 144 & 185.244(3) & 0.000 & $^3D$ & 3 & 455612.2 &  & $(^1S)^1G$  & 4 & 995441 &              &             \\
185.978 &          & 450 & 219 & 185.978(3) & 0.000 & $^3F$ & 2 & 433818.8 &  & $(^3P)^3Fb$ & 2 & 971516 & 185.994 YLTW &             \\
187.950 &          & 480 & 1461 & 187.950(3) & 0.000 & $^1P$ & 1 & 584546 &  & $(^3P)^1D a$ & 2 & 1116602 & 187.961 YLTW &             \\
188.300 & bl(XI)   & 1000 & 149 & 188.303(2) & -0.003 & $^3F$ & 2 & 433818.8 &  & $(^1D)^1F a$ & 3 & 964878 &              &             \\
188.493 &          & 1920 & 1291 & 188.493(4) &        & $^3F$ & 4 & 425809.8 &  & $(^3P)^3Gb$ & 5 & 956333 & 188.497 Y09  & 188.493 D09 \\
188.678 &          & 240 & 79 & 188.680(3) & -0.002 & $^3F$ & 4 & 425809.8 &  & $(^3P)^3Gb$ & 4 & 955807 & 188.685 YL   &             \\
188.818 &          & 1160 & 745 & 188.817(3) & 0.001 & $^1F$ & 3 & 465828.4 &  & $(^1S)^1G$  & 4 & 995441 & 188.823 LYTW & 188.807 D09 \\
189.577 &          & 400 & 130 & 189.578(3) & -0.001 & $^3F$ & 3 & 429310.9 &  & $(^3P)^3Gb$ & 3 & 956798 & 189.582 YL   &             \\
189.937 &          & 1820 & 958 & 189.935(3) & 0.002 & $^3F$ & 3 & 429310.9 &  & $(^3P)^3Gb$ & 4 & 955807 & 189.941 Y09  & 189.935 D09 \\
191.210 &          & 1290 & 544 & 191.212(3) & -0.002 & $^3F$ & 2 & 433818.8 &  & $(^3P)^3Gb$ & 3 & 956798 & 191.216 Y09  & 191.206 D09 \\
192.182 &          & 200 & 28 & 192.174(2) & 0.008 & $^3D$ & 3 & 455612.2 &  & $(^3P)^3Fb$ & 3 & 975974 &              &             \\
192.603 &          & 100 & 33 & 192.596(3) & 0.007 & $^1D$ & 2 & 456752.7 &  & $(^3P)^3Fb$ & 3 & 975974 &              &             \\
192.633 & bl(XI)   & 1040 & 374 & 192.630(3) & 0.003 & $^3D$ & 3 & 455612.2 &  & $(^3P)^3Fb$ & 4 & 974743 & 192.632 YL & 192.630 D14 \\
194.796 &          & 940 & 302 & 194.796(2) & 0.000 & $^3D$ & 2 & 462616.6 &  & $(^3P)^3Fb$ & 3 & 975974 & 194.806 YLTW & 194.784\tablenotemark{h} D09 \\
195.733 &          & 670 & 209 & 195.733(3) & 0.000 & $^3D$ & 1 & 460616 &  & $(^3P)^3Fb$ & 2 & 971516 & 195.743 YLTW & 195.743 D09 \\
196.356 &          & 300 & 35 & 196.361(2) & -0.005 & $^3D$ & 3 & 455612.2 &  & $(^1D)^1F a$ & 3 &  964878&               &             \\
196.499 &          & 190 & 39 & 196.496(3) & 0.003 & $^1F$ & 3 & 465828.4 &  & $(^3P)^3Fb$ & 4 & 974743 &              &             \\
196.805 &          & 530 & 338 & 196.802(2) & 0.003 & $^1D$ & 2 & 456752.7 &  & $(^1D)^1F a$ & 3 & 964878 & 196.810 YLTW & 196.803 D09 \\
199.985 &          & 360 & 29 & 199.982(3) & 0.003 & $^1D$ & 2 & 456752.7 &  & $(^3P)^3Gb$ & 3 & 956798 & 199.986 YL   &             \\
200.386 &          & 200 & 72 & 200.381(2) & 0.005 & $^1F$ & 3 & 465828.4 &  & $(^1D)^1F a$ & 3 & 964878 &              &             \\
\enddata
\tablenotetext{a}{Character of the observed line: db -- intensity is shared by two transitions; bl -- blended line (the blending species are given in parentheses where known): VII -- \ion{Fe}{vii}, VIII -- \ion{Fe}{viii}, XI -- \ion{Fe}{xi}; ? -- identification is uncertain; p -- perturbed by a stronger nearby line (both the wavelength and intensity may be affected); w -- wide line.}
\tablenotetext{b}{Relative intensities are given on an arbitrary linear scale (see text).}
\tablenotetext{c}{Weighted transition probability ($g$ is the statistical weight of the upper level) in $10^9$~s$^{-1}$ unit.}
\tablenotetext{d}{Wavelength derived from the final level energies (Ritz wavelength).}
\tablenotetext{e}{Difference between the observed and Ritz wavelengths (blank for lines that solely determine the upper level).}
\tablenotetext{f}{Designation is restricted to a term of the $3p^4$ sub-shell followed by a final term and a letter (a or b) distinguishing different terms of the $3d^2$ configuration in the case the final term is repeated. For full designation see Table~\ref{tbl.4}.}
\tablenotetext{g}{Previous measurements and identifications: BT18TW – measured by \citet{2018ApJ...854..114B}, identification of this work; D09 -- \citet{2009A+A...508..501D}; D14 -- \citet{2014A+A...565A..77D}; Y09 -- \citet{2009ApJ...691L..77Y}; LYTW -- line from \citet{2009ApJ...706....1L}, identification of this work; YL -- \citet{2009ApJ...707..173Y}; YLTW -- suggested as \ion{Fe}{ix} lines by \citet{2009ApJ...707..173Y}, identification of this work.}
\tablenotetext{h}{Lines are not fully resolved from, respectively, \ion{Fe}{vii} 176.905 and 194.770~\AA\ lines in EIS spectra.}
\tablenotetext{i}{Line is blended with \ion{O}{v} line in EBIT spectrum.}
\end{deluxetable}

\pagebreak

\startlongtable
\begin{deluxetable}{cccccccc}
\tablecaption{Energy levels of the $3p^43d^2$ configuration of \ion{Fe}{ix} higher than 930\,000~cm$^{-1}$.\label{tbl.4}}
\tablehead{
   \colhead{$E$ (cm$^{-1}$)\tablenotemark{a}} &
   \colhead{o--c\tablenotemark{b}} & 
   \colhead{$J$} &
   \multicolumn4c{Composition\tablenotemark{c}} & 
   \colhead{N\tablenotemark{d}}
}
\startdata
 931158\#  &      & 1 & 61\%\ $(^3P)^3Da$ & [$3p^4(^3P)3d^2(^3F)^3D$] & 20\%\ $(^1D)^3Da$ &    9\%\ $(^2S)^3D$* &  \\
 934730\#  &      & 2 & 56\%\ $(^3P)^3Da$ & [$3p^4(^3P)3d^2(^3F)^3D$] & 18\%\ $(^1D)^3Da$ &    8\%\ $(^2S)^3D$* &  \\
 934973\#  &      & 1 & 74\%\ $(^3P)^3P$  & [$3p^4(^3P)3d^2(^1S)^3P$] & 12\%\ $(^3P)^3Pa$ &    6\%\ $(^3P)^3Pb$ &  \\
 937428\# &  & 5 & 96\%\ $(^1D)1H$ & [$3p^4(^1D)3d^2(^1G)1H$] &  6\%\ $(^1D)^3G$ &    1\%\ $(^3P)3H$ &  \\
 937854\#  &      & 0 & 73\%\ $(^3P)^3P$  & [$3p^4(^3P)3d^2(^1S)^3P$] & 12\%\ $(^3P)^3Pa$ &    6\%\ $(^3P)^3Pb$ &  \\
 939790\#  &      & 3 & 56\%\ $(^3P)^3Da$ & [$3p^4(^3P)3d^2(^3F)^3D$] & 20\%\ $(^1D)^3Da$ &    8\%\ $(^2S)^3D$* &  \\
 943984\# &       & 2 & 55\%\ $(^1D)^3Fb$ & [$3p^4(^1D)3d^2(^3P)^3F$] & 28\%\ $(^1S)^3F$  &  5\%\ $(^1D)^3Fa$ &  \\
 946522\# &       & 3 & 45\%\ $(^1D)^3Fb$ & [$3p^4(^1D)3d^2(^3P)^3F$] & 28\%\ $(^1S)^3F$  &  6\%\ $(^1D)^3Fa$ &  \\
 949555\# &       & 4 & 44\%\ $(^1D)^3Fb$ & [$3p^4(^1D)3d^2(^3P)^3F$] & 38\%\ $(^1S)^3F$  &  6\%\ $(^1D)^3Fa$ &  \\
 955807(9) & -279 & 4 & 37\%\ $(^3P)^3Gb$ & [$3p^4(^3P)3d^2(^1G)^3G$] & 32\%\ $(^1D)^3G$  & 23\%\ $(^3P)^3Ga$ & 2 \\
 956333(11) & 13 & 5 & 41\%\ $(^3P)^3Gb$ & [$3p^4(^3P)3d^2(^1G)^3G$] & 33\%\ $(^1D)^3G$  & 22\%\ $(^3P)^3Ga$ & 1 \\
 956798(3) & -395 & 3 & 25\%\ $(^3P)^3Gb$ & [$3p^4(^3P)3d^2(^1G)^3G$] & 25\%\ $(^1D)^1Fa$ & 23\%\ $(^1D)^3G$ & 3 \\
 964878(10) & -102 & 3 & 64\%\ $(^1D)^1Fa$ & [$3p^4(^1D)3d^2(^1G)^1F$] & 14\%\ $(^3P)^3Gb$ &  9\%\ $(^1D)^3G$ & 4 \\
 971516(8) & -42 & 2 & 41\%\ $(^3P)^3Fb$ & [$3p^4(^3P)3d^2(^1G)^3F$] & 38\%\ $(^1D)^3Fa$ & 15\%\ $(^1S)^3F$ & 2 \\
 974743(11) & -78 & 4 & 37\%\ $(^3P)^3Fb$ & [$3p^4(^3P)3d^2(^1G)^3F$] & 33\%\ $(^1D)^3Fa$ & 12\%\ $(^1S)^3F$ & 3 \\
 975974(12) & 283 & 3 & 39\%\ $(^3P)^3Fb$ & [$3p^4(^3P)3d^2(^1G)^3F$] & 33\%\ $(^1D)^3Fa$ & 12\%\ $(^1S)^3F$ & 5 \\
 982218\# &       & 0 & 55\%\ $(^1S)^3P$  & [$3p^4(^1S)3d^2(^3P)^3P$] & 40\%\ $(^1D)^3Pb$ &  2\%\ $(^1D)^3Pa$ &  \\
 982956(12) & -179 & 1 & 79\%\ $(^3P)^3S$  & [$3p^4(^3P)3d^2(^3P)^3S$] & 13\%\ $(^1S)^3P$  &  2\%\ $(^2P)^3S$ & 3 \\
 983770\# &       & 1 & 47\%\ $(^1S)^3P$  & [$3p^4(^1S)3d^2(^3P)^3P$] & 33\%\ $(^1D)^3Pb$ &  9\%\ $(^3P)^3S$ &  \\
 983927\# &       & 2 & 32\%\ $(^1D)^1D$  & [$3p^4(^1D)3d^2(^1S)^1D$] & 23\%\ $(^1S)^3P$  & 13\%\ $(^1S)^1D$ &  \\
 987829\# &       & 2 & 38\%\ $(^1S)^3P$  & [$3p^4(^1S)3d^2(^3P)^3P$] & 17\%\ $(^1D)^1D$  & 17\%\ $(^1D)^3Pb$ &  \\
 990852(13) & 843 & 3 & 47\%\ $(^1D)^3Db$ & [$3p^4(^1D)3d^2(^3P)^3D$] & 14\%\ $(^3P)^3Db$ & 13\%\ $(^3P)^3Da$ & 1 \\
 991610\# &       & 4 & 58\%\ $(^1D)^1Ga$ & [$3p^4(^1D)3d^2(^1G)^1G$] & 26\%\ $(^1S)^1G$  & 11\%\ $(^3P)^1G$ &  \\
 992387(9) & 486 & 2 & 38\%\ $(^1D)^3Db$ & [$3p^4(^1D)3d^2(^3P)^3D$] & 13\%\ $(^3P)^3Db$ & 12\%\ $(^3P)^3Da$ & 3 \\
 993425(13) & -102 & 1 & 41\%\ $(^1D)^3Db$ & [$3p^4(^1D)3d^2(^3P)^3D$] & 14\%\ $(^3P)^3Db$ & 12\%\ $(^3P)^3Da$ & 2 \\
 993490(14) & 125 & 0 & 48\%\ $(^1D)^3Pa$ & [$3p^4(^1D)3d^2(^3F)^3P$] & 35\%\ $(^3P)^3Pb$ & 10\%\ $(^1S)^3P$ & 1 \\
 995441(7) & -254 & 4 & 51\%\ $(^1S)^1G$  & [$3p^4(^1S)3d^2(^1G)^1G$] & 25\%\ $(^3P)^1G$  & 10\%\ $(^1D)^1Ga$ & 3 \\
996057(10) & 732 & 1 & 41\%\ $(^1D)^3Pa$ & [$3p^4(^1D)3d^2(^3F)^3P$] & 31\%\ $(^3P)^3Pb$ & 10\%\ $(^1S)^3P$ & 4 \\
996147(10) & -497 & 2 & 38\%\ $(^1D)^3Pa$ & [$3p^4(^1D)3d^2(^3F)^3P$] & 35\%\ $(^3P)^3Pb$ & 15\%\ $(^1S)^3P$ & 2 \\
1014356(8) & 172 & 4 & 48\%\ $(^3P)^3Fa$ & [$3p^4(^3P)3d^2(^3F)^3F$] & 24\%\ $(^1S)^3F$  & 11\%\ $(^3P)^3F$ & 4 \\
1017204(11) & 245 & 2 & 32\%\ $(^3P)^1Db$ & [$3p^4(^3P)3d^2(^3P)^1D$] & 18\%\ $(^1D)^1Db$ & 15\%\ $(^1D)^1D$ & 3 \\
1018894(9) & 204 & 3 & 44\%\ $(^3P)^3Fa$ & [$3p^4(^3P)3d^2(^3F)^3F$] & 23\%\ $(^1S)^3F$  & 11\%\ $(^3P)^3F$ & 5 \\
1022856(7) & 582 & 2 & 27\%\ $(^3P)^3Fa$ & [$3p^4(^3P)3d^2(^3F)^3F$] & 15\%\ $(^1S)^3F$  &  7\%\ $(^1D)^1Da$ & 6 \\
1023007(13)? & -220 & 1 & 42\%\ $(^3P)^1P$  & [$3p^4(^3P)3d^2(^3P)^1P$] & 25\%\ $(^1D)^1P$  &  9\%\ $(^3P)^3Pa$ & 1 \\
1023390\# &       & 0 & 50\%\ $(^1S)^1S$  & [$3p^4(^1S)3d^2(^1S)^1S$] & 14\%\ $(^3P)^3Pa$ & 13\%\ $(^1D)^1S$ &  \\
1024953(10) & -331 & 2 & 31\%\ $(^3P)^3Pa$ & [$3p^4(^3P)3d^2(^3P)^3P$] & 18\%\ $(^1D)^3Pb$ & 12\%\ $(^3P)^3P$ & 3 \\
1028110(9) & -388 & 2 & 22\%\ $(^1D)^1Da$ & [$3p^4(^1D)3d^2(^1G)^1D$] & 15\%\ $(^2S)^1D$* & 12\%\ $(^3P)^3Fa$ & 3 \\
1029528(13)? & -1504 & 0 & 23\%\ $(^1S)^1S$  & [$3p^4(^1S)3d^2(^1S)^1S$] & 21\%\ $(^3P)^3Pa$ & 15\%\ $(^3P)^3P$ & 1 \\
1029709(17) & 317 & 1 & 23\%\ $(^3P)^3Pa$ & [$3p^4(^3P)3d^2(^3P)^3P$] & 14\%\ $(^3P)^1P$  & 10\%\ $(^1S)^3P$ & 2 \\
1030255(15) & -1093 & 3 & 34\%\ $(^1D)^3Da$ & [$3p^4(^1D)3d^2(^3F)^3D$] & 20\%\ $(^3P)^3Db$ & 17\%\ $(^3P)^3Da$ & 2 \\
1035659(12) & -1451 & 1 & 36\%\ $(^1D)^3Da$ & [$3p^4(^1D)3d^2(^3F)^3D$] & 24\%\ $(^3P)^3Db$ & 18\%\ $(^3P)^3Da$ & 2 \\
1037191(7) & 197 & 2 & 31\%\ $(^1D)^3Da$ & [$3p^4(^1D)3d^2(^3F)^3D$] & 21\%\ $(^3P)^3Db$ & 17\%\ $(^3P)^3Da$ & 6 \\
1046472(10) & 1158 & 3 & 54\%\ $(^3P)^1F$  & [$3p^4(^3P)3d^2(^3F)^1F$] & 19\%\ $(^1D)^1Fb$ & 19\%\ $(^2P)^3D$** & 4 \\
1116602(8) & 8 & 2 & 54\%\ $(^3P)^1Da$ & [$3p^4(^3P)3d^2(^3F)^1D$] & 24\%\ $(^1D)^1Da$ &  7\%\ $(^3P)^1Db$ & 3 \\
1188998(15)? & 136 & 0 & 63\%\ $(^3P)^1S$  & [$3p^4(^3P)3d^2(^3P)^1S$] &  17\%\ $(^1D)^1S$  & 12\%\ $(^1S)^1S$ & 1 \\
\enddata
\tablenotetext{a}{\# -- calculated value for the level; ? -- questionable levels not included in the fitting.}
\tablenotetext{b}{Residual (``obs.--calc.") of the parametric least-squares fit with Cowan's codes in cm$^{-1}$.}
\tablenotetext{c}{$LS$ -- composition of the level eigenvector. The $3p^4$ configuration terms and the final terms are listed. The letters after the final terms distinguish different terms of the $3d^2$ configuration. Full descriptions for the first component are shown in square brackets. * -- $(^2S)^1D$ and $(^2S)^3D$ terms belong to the $3s3p^63d$ configuration; ** - $(^2P)^3D$ stands for $3p^5(^2P)4p(^2P)^3D$.}
\tablenotetext{d}{Number of observed lines determining the level value in the least-squares optimization procedure.}
\end{deluxetable}

\pagebreak

\startlongtable
\begin{deluxetable}{llrrrr}
\tablecaption{Hartree–Fock with relativistic corrections (HFR) and least-square-fitted (LSF) parameter values (cm$^{-1}$) with their uncertainties (Unc.) in \ion{Fe}{ix}.\label{tbl.5}}
\tablehead{
  \colhead{Configuration} &
  \colhead{Parameter} &
  \colhead{LSF} &
  \colhead{Unc.\tablenotemark{a}} &
  \colhead{HFR\tablenotemark{b}} &
  \colhead{LSF/HFR\tablenotemark{c}}
}
\startdata
\sidehead{Even configurations}
$3s^23p^6$      & $E_\mathrm{av}$        & 31763 & 571 & 30744 & 1.032\\
\noalign{\smallskip}
$3s^23p^54p$   & $E_\mathrm{av}$        & 1054771 &   f  & 1084771 & 0.971\\
\noalign{\smallskip}
$3s^23p^55p$   & $E_\mathrm{av}$        & 1406545 &   f  & 1436545 & 0.978\\
\noalign{\smallskip}
$3s^23p^56p$   & $E_\mathrm{av}$        & 1573258 &   f  & 1603258 & 0.980\\
\noalign{\smallskip}
$3s^23p^54f$    & $E_\mathrm{av}$        & 1310977 & 165 & 1340635 & 0.977\\
            & $\zeta$(3p)   & 10340 & 330 & 10130 & 1.021\\
            & $\zeta$(4f)   & 21 &   f  & 21 & 1.000\\
            & $F^2(3p, 4f)$ & 36124 & 2265 & 38697 & 0.933\\
            & $G^2(3p, 4f)$ & 18411 & 1313 & 21281 & 0.865\\
            & $G^4(3p, 4f)$ & 13303 & 4016 & 14048 & 0.947\\
\noalign{\smallskip}
$3s^23p^55f$    & $E_\mathrm{av}$        & 1520529 &   f  & 1550529 & 0.980\\
\noalign{\smallskip}
$3s^23p^56f$    & $E_\mathrm{av}$        & 1634436 &   f  & 1664436 & 0.981\\
\noalign{\smallskip}
$3s^23p^43d^2$   & $E_\mathrm{av}$       & 906997 & 463 & 938864 & 0.964\\
            & $F^2(3p,3p)$ & 127124 & 3697 & 132065 & 0.963\\
            & $\alpha$(3p)   & 199 & 223 &         & \\
            & $F^2(3d,3d)$ & 105070 & 3063 & 131852 & 0.797\\
            & $F^4(3d,3d)$ & 63148 & 4848 & 84189 & 0.750\\
            & $\zeta(3p)$   & 10012 & 482 & 9698 & 1.032\\
            & $\zeta(3d)$   & 753 & 169 & 798 & 0.944\\
            & $F^2(3p,3d)$ & 117806 & 1170 & 128513 & 0.917\\
            & $G^1(3p,3d)$ & 128628 & 598 & 153097 & 0.840\\
            & $G^3(3p,3d)$ & 79137 & 2044 & 97017 & 0.816\\
\noalign{\smallskip}
$3s3p^63d$     & $E_\mathrm{av}$        & 800263 & 977 & 829259 & 0.962\\
            & $\zeta(3d)$   & 755 & 169 & 800 & 0.944\\
            & $G^2(3s,3d)$ & 120637 & 2520 & 119177 & 1.012\\
\noalign{\smallskip}
$3s3p^64s$     & $E_\mathrm{av}$        & 1314892 &   f  & 1344892 & 0.976\\
            & $G^0(3s, 4s)$ & 8928 &   f  & 10505 & 0.850\\
\noalign{\smallskip}
$3s^23p^43d4s$  & $E_\mathrm{av}$        & 1406139 &   f  & 1436139 & 0.978\\
\noalign{\smallskip}
$3s^23p^44s2$   & $E_\mathrm{av}$        & 1975829 &   f  & 2005829 & 0.984\\
\noalign{\smallskip}
$3s^23p^44s4d$  & $E_\mathrm{av}$        & 2220346 &   f  & 2250346 & 0.986\\
 & $R^1(3s,3d;3p,3p)$\tablenotemark{d} & 145004 & 897 & 160894 & 0.901\\
            & ${\sigma}$\tablenotemark{e} & 555 &      &         & \\
\noalign{\smallskip}
\sidehead{Odd configurations}
$3s^23p^53d$   & $E_\mathrm{av}$        & 448456 & 16 & 477759 & 0.932\\
            & $\zeta(3p)$   & 9921 & 48 & 9686 & 1.024\\
            & $\zeta(3d)$   & 784 & 18 & 798 & 0.982\\
            & $F^1(3p,3d)$ & 4862 & 121 & 0 & \\
            & $F^2(3p,3d)$ & 115238 & 143 & 128765 & 0.895\\
            & $G^1(3p,3d)$ & 129360 & 47 & 153302 & 0.844\\
            & $G^2(3p,3d)$ & 5252 & 245 & 0 & \\
            & $G^3(3p,3d)$ & 84578 & 258 & 97191 & 0.870\\
\noalign{\smallskip}
$3s^23p^54d$   & $E_\mathrm{av}$        & 1199786 & 38 & 1233059 & 0.971\\
            & $\zeta(3p)$   & 10616 & 50 & 10082 & 1.053\\
 & $\zeta(4d)$ & 220 &   f  & 220 & 1.000\\
 & $F^2(3p, 4d)$ & 31127 &   f & 36620 & 0.850\\
 & $G^1(3p, 4d)$ & 8573 &   f & 10085 & 0.850\\
 & $G^3(3p, 4d)$ & 8520 &   f & 10023 & 0.850\\
\noalign{\smallskip}
$3s^23p^55d$   & $E_\mathrm{av}$        & 1469453 &   f  & 1502453 & 0.977\\
\noalign{\smallskip}
$3s^23p^56d$   & $E_\mathrm{av}$        & 1605451 &   f  & 1638451 & 0.979\\
\noalign{\smallskip}
$3s^23p^54s$    & $E_\mathrm{av}$        & 954374 & 39 & 985623 & 0.966\\
            & $\zeta(3p)$   & 10307 & 58 & 10077 & 1.023\\
            & $G^1(3p,4s)$ & 11655 &   f  & 13711 & 0.850\\
\noalign{\smallskip}
$3s^23p^55s$    & $E_\mathrm{av}$        & 1362553 & 38 & 1392738 & 0.977\\
            & $\zeta(3p)$   & 10292 & 49 & 10138 & 1.015\\
            & $G^1(3p,5s)$ & 3940 &   f  & 4636 & 0.850\\
\noalign{\smallskip}
$3s3p^64p$     & $E_\mathrm{av}$        & 1408803 &   f  & 1440803 & 0.977\\
\noalign{\smallskip}
$3s3p^64f$     & $E_\mathrm{av}$        & 1665072 &   f  & 1697072 & 0.980\\
\noalign{\smallskip}
 & $\sigma$\tablenotemark{e} & 51 &      &         & \\
\enddata
\tablenotetext{a}{f -- fixed parameter.}
\tablenotetext{b}{Average energies are adjusted so that the energy of the ground level $3s^23p^6$ $^1S_0$ is zero in the calculation of the even configuration matrix with all electrostatic parameters scaled by 0.85 factor.}
\tablenotetext{c}{Electrostatic parameters of the configurations, as well as interaction parameters not listed in the table, are scaled by a factor of 0.85 with respect to ab initio values; the spin-orbit parameters are not scaled.}
\tablenotetext{d}{Parameter of interaction between the $3s^23p^43d^2$ and $3s3p^63d$ configurations.}
\tablenotetext{e}{Root mean square deviation of the fitting.}
\end{deluxetable}



\end{document}